\documentclass[12pt]{iopart}

\usepackage{iopams}
\usepackage{setstack}
\usepackage{subfig}
\usepackage{graphicx}
\usepackage{float}
\usepackage{cite}
\usepackage[labelfont=bf, textfont=it]{caption}
\usepackage{hyperref}
\usepackage{xcolor}
\usepackage{tabularx}

\begin{document}

\title{Nonlinear microtearing modes in MAST and their stochastic layer formation}

\author{M. Giacomin$^1$, D. Dickinson$^1$, D. Kennedy$^2$, B. S. Patel$^2$ and C. M. Roach$^2$}

\address{$^1$York Plasma Institute, University of York, York, YO10 5DD, United Kingdom}
\address{$^2$Culham Centre for Fusion Energy, Abingdon OX14 3DB, United Kingdom}
\ead{maurizio.giacomin@york.ac.uk}

\begin{abstract}
First nonlinear gyrokinetic simulations of microtearing modes in the core of a MAST case are performed on two surfaces of the high-collisionality discharge used in Valovič \emph{et al.} Nucl. Fusion 51.7 (2011) to obtain the favorable energy confinement scaling with collisionality, $\tau_E\propto\,\nu_*^{-1}$. On the considered surfaces microtearing modes dominate linearly at binormal length scales of the order of the ion Larmor radius.
While the effect of electron collision frequency is moderate in linear simulations, a strong dependence on this parameter is found in nonlinear simulations at $r/a=0.5$, where $r$ and $a$ are the surface and tokamak minor radius, respectively.
The dynamics of magnetic islands generated by microtearing modes is analysed, showing that the radial extent of the stochastic region caused by islands overlapping plays an important role in determining the saturation level of the microtearing mode driven heat flux.
Local nonlinear gyrokinetic simulations show that the microtearing mode driven heat flux, $Q_e^\mathrm{MTM}$, is largely dominated by magnetic flutter and depends strongly on the magnetic shear, $\hat{s}$. Comparing two surfaces, $r/a=0.5$ and $r/a=0.6$, reveals that $Q_e^\mathrm{MTM}$ is negligible at $r/a=0.5$ ($\hat{s}=0.34$), with the electron temperature gradient driven heat flux, $Q_e^\mathrm{ETG}$, comparable to the experimental electron heat flux, $Q_e^\mathrm{exp}$, while $Q_e^\mathrm{MTM}$ is significantly larger and comparable to $Q_e^\mathrm{ETG}$ and $Q_e^\mathrm{exp}$ at $r/a=0.6$ ($\hat{s}=1.1$). Microtearing modes cause more experimentally significant transport in higher $\hat{s}$ regions and may influence (together with electron temperature gradient modes) the observed scaling of energy confinement time with collisionality (Valovič \emph{et al.} Nucl. Fusion 51.7 (2011)).

\end{abstract}

%

\section{Introduction}

Previous theoretical and numerical works have shown the presence of a linear tearing instability at high mode numbers driven by an electron temperature gradient and denoted as microtearing mode (MTM)~\cite{hazeltine1975}. Recent nonlinear simulations have shown that the MTM instability can significantly contribute to the electron heat flux in the edge of H-mode plasmas as well as in the core of spherical tokamaks (see, e.g., Refs.~\cite{guttenfelder2011,doerk2011,hatch2016,maeyama2017,pueschel2020}).
The first theoretical description of the MTM instability has been proposed in Ref.~\cite{hazeltine1975} and extended later in Refs.~\cite{drake1977,gladd1980,hassam1980,catto1981,cowley1986,connor1990}.
In particular, Ref.~\cite{drake1977} shows that the MTM instability separates into three regimes depending on the electron collision frequency, $\nu_e=4\pi n_e e^4\ln\lambda/[(2T_e)^{3/2}m_e^{1/2}]$ ($n_e$ is the electron density, $\lambda$ is the Coulomb logarithm, $T_e$ is the electron temperature and $m_e$ is the electron mass): a collisionless regime with $\nu_e \ll \omega$, a semi-collisional regime with $\omega \sim (k_\parallel v_{\mathrm{th, e}})^2/\nu_e < \nu_e$  (with $v_\mathrm{th, e} = \sqrt{2T_e/m_e}$ the electron thermal velocity and $k_\parallel$ the parallel wave vector), and a collisional regime with $\nu_e \gg \omega$, where $\omega$ is the MTM frequency.
The work in Ref.~\cite{drake1977} neglects the effect of the electrostatic potential in the collisionless and semi-collisional regimes. A following numerical analysis has extended this work by including the effect of electrostatic potential fluctuations, showing that these provide a strong destabilising effect~\cite{gladd1980}, also confirmed in recent linear gyrokinetic simulations~\cite{hamed2019}. 
While first linear studies show that the mechanism driving the collisional MTM instability requires a velocity dependent collision frequency~\cite{hassam1980}, unstable (collisional) MTMs have been found also when a velocity independent collision operator is considered~\cite{applegate2007}, highlighting the presence of various driving mechanisms. 

A magnetic perturbation $\delta B_{mn}$ associated with MTMs resonates at the rational surface with $q=m/n$, where $q$ is the safety factor, $m$ and $n$ are the poloidal and toroidal magnetic perturbation mode number, respectively. 
Resonant modes can reconnect and form magnetic islands. An estimate of the island width is derived in Ref.~\cite{wesson2011},  
\begin{eqnarray}
w_\mathrm{island} = 4 \sqrt{\frac{\delta B}{B_0} \frac{rR}{n\hat{s}}}\,,
\label{eqn:wisland}
\end{eqnarray}
where $B_0$ is the unperturbed magnetic field, $r$ the tokamak minor radius, $R$ the tokamak major radius and $\hat{s}=(r/q)\mathrm{d}q/\mathrm{d}r$ is the magnetic shear.
The distance between two rational surfaces with consecutive $m$ and same $n$, corresponding to $q(r_m) = m/n$ and $q(r_{m+1}) = (m+1)/n$, is approximated by 
\begin{equation}
\label{eqn:sep_single}
\Delta r \simeq 1/(nq') = r/(nq\hat{s})\,,    
\end{equation}
where $q'=\mathrm{d}q/\mathrm{d}r$. 
In a typical flux-tube calculation resolving toroidal mode numbers $\{n_0, 2n_0,\dots, n=N n_0\}$ the minimum spacing between rational surfaces is given by~\cite{nevins2011}
\begin{eqnarray}
\delta r \simeq \frac{n_0r}{n^2 q\hat{s}}\,.
\label{eqn:sep}
\end{eqnarray} 
If the magnetic island width is larger than the distance between two adjacent rational surfaces, a region of stochastic magnetic field lines can form~\cite{martin1984}.
Magnetic field stochasticity can provide a strong transport mechanism, as described in Ref.~\cite{rechester1978}, thus accounting for the significant electron heat flux observed in nonlinear gyrokinetic MTM simulations~\cite{guttenfelder2011}. In addition, electron heat transport consistent with the island overlap criterion has been observed in NSTX experiments~\cite{wong2008}. 

While linear gyrokinetic simulations have been extensively carried out and show the presence of MTMs in many experimentally relevant scenarios~\cite{roach2005,applegate2007,told2008,valovivc2011,dickinson2013,moradi2013,hatch2016,patel2021} and particularly in spherical tokamaks~\cite{kaye2021}, the evaluation of the electron heat flux driven by MTMs requires one to perform nonlinear gyrokinetic simulations, which remain very challenging because of the high numerical requirements~\cite{doerk2011}.
In recent years, significant effort has been devoted to understand the mechanisms behind the saturation of the MTM driven electron heat flux.
For example, Ref.~\cite{guttenfelder2011} shows that the MTM driven heat flux can be significantly reduced by equilibrium flow shear.
Zonal fields~\cite{pueschel2020} and local temperature flattening~\cite{ajay2022} have also been linked to the saturation of MTM turbulence. In Ref.~\cite{maeyama2017}, ion-scale MTMs are suppressed by electron-scale turbulence via cross-scale nonlinear interactions.

Refs.~\cite{guttenfelder2011,guttenfelder2012,guttenfelder2012b} shows that MTMs can drive significant electron heat transport in NSTX and links the collisionality dependence of the energy confinement time observed in NSTX to the collisionality dependence of MTM-driven heat flux. 
Analogously, Ref.~\cite{valovivc2011} suggests a similar role played by MTM turbulence in MAST, in particular noting that, if MTMs were to dominate heat transport, lowering the electron collisionality would reduce the electron heat flux from MTMs and would be consistent with the observed energy confinement scaling $\tau_e \propto 1/\nu_*$. 
While gyrokinetic nonlinear simulations of MTMs have been performed in NSTX to support this hypothesis~\cite{guttenfelder2012b}, only gyrokinetic linear simulations have been carried out in Ref.~\cite{valovivc2011}.

In this work, we extend the linear study presented in Ref.~\cite{valovivc2011} by performing a set of nonlinear gyrokinetic simulations of experimentally relevant cases built from the MAST discharge \#22769~\cite{valovivc2011}: the simulations reported here are the first converged nonlinear simulations of MTM turbulence in MAST.
All the cases considered here are characterised by a dominant ion scale collisional MTM instability, whose dependence on various parameters and, in particular, on the electron collision frequency is investigated. An electron temperature gradient (ETG) instability is found at electron scale.
We also find that the MTM-driven heat flux $Q_e^{\mathrm{MTM}}$ is sensitive to magnetic shear and to electron collision frequency, $\nu_e$.
Local gyrokinetic simulations on two neighbouring surfaces in this MAST plasma at $r/a=0.5$ and $r/a=0.6$ reveal that $Q_e^\mathrm{MTM}$ is negligible at $r/a=0.5$ (lower magnetic shear, $\hat{s}=0.34$), with $Q_e^\mathrm{ETG}$ comparable to $Q_e^\mathrm{exp}$, while $Q_e^\mathrm{MTM}$ is substantial and comparable to both $Q_e^\mathrm{exp}$  and $Q_e^\mathrm{ETG}$ at $r/a=0.6$ (higher magnetic shear, $\hat{s}=1.1$). 

In addition, this MAST equilibrium provides a useful reference for a scientific study of nonlinear saturation of MTMs in numerically tractable conditions.
Saturation mechanisms are investigated, and indicate a significant contribution from zonal fields in the saturation process, in agreement with Ref.~\cite{pueschel2020}.
The level of saturated heat flux is found to strongly depend on the radial extent of the stochastic region due to magnetic island overlapping.

This paper is organised as follows. The MAST reference case is introduced in Sec.~\ref{sec:reference}, where the dominant microinstabilities are identified by means of linear gyrokinetic simulations. 
In Sec.~\ref{sec:linear}, the main MTM instability is characterised and results from linear simulation scans in electron temperature gradient, density gradient and electron collision frequency are presented. 
Results of nonlinear simulations at $r/a=0.5$ are reported in Sec.~\ref{sec:nonlinear}, where the main saturation mechanism is identified. The effect of the stochastic layer formation on heat flux is analysed in Sec.~\ref{sec:islands}. The results of linear and nonlinear simulations carried out on the nearby $r/a=0.6$ flux surface with higher magnetic shear are discussed in Sec.~\ref{sec:rad06}.  Conclusions follow in Sec.~\ref{sec:conclusion}.

\section{The MAST reference case}
\label{sec:reference}

\begin{figure}
    \centering
    \subfloat[]{\includegraphics[width=0.35\textwidth]{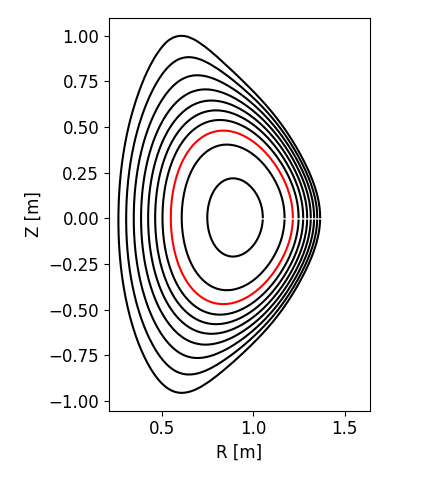}}\quad
    \subfloat[]{\includegraphics[width=0.55\textwidth]{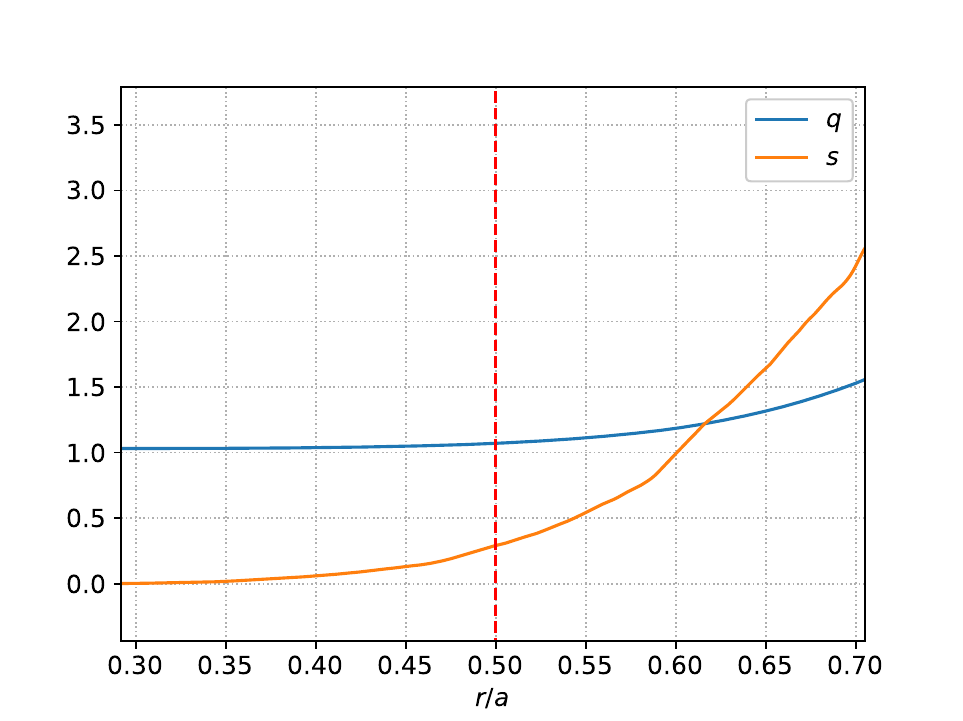}}\quad
    \caption{(a) Contour plot of the equilibrium poloidal magnetic flux for the MAST discharge \#22769 at $t=0.2$~s~\cite{valovivc2011}. The red line corresponds to the reference radial surface located at $r/a=0.5$. (b) Safety factor and magnetic shear profile in the region $r/a\in [0.3, 0.7]$ around the relevant radial surfaces. The red dashed vertical line indicates the radial position of the reference surface at $r/a=0.5$.}
    \label{fig:equilibrium}
\end{figure}

The reference case is based on the MAST discharge \#22769, which corresponds to the high collisionality discharge from a two point collisionality scan in MAST, where the energy confinement time was found to scale approximately as $\tau_e\propto \nu_*^{-0.8}$~\cite{valovivc2011}, where $\nu_*=\nu_e qR/(\epsilon^{3/2}v_{\mathrm{th,e}})$.
The equilibrium and the kinetic profiles used here have been obtained by running TRANSP.
Equilibrium magnetic flux surfaces for this discharge are shown in Fig.~\ref{fig:equilibrium} at $t=0.2$~s.
A linear gyrokinetic analysis carried out in Ref.~\cite{valovivc2011} shows that MTMs are the dominant linear microinstability at $k_y\rho_s\lesssim 1$ ($\rho_s=c_s/\Omega_{i}$ is the ion sound Larmor radius with $c_s=\sqrt{T_e/m_D}$ the ion sound speed, $\Omega_i=eB/m_D$ the ion cyclotron frequency and $m_D$ the deuterium mass), thus suggesting a possible important role played by MTM turbulence in this discharge (only linear simulations were performed in Ref.~\cite{valovivc2011}).
Therefore this experimental case is of particular interest to characterise and analyse MTM turbulence and transport and is considered as a baseline case for the following analysis.
Details on the equilibrium and profiles are reported in Ref.~\cite{valovivc2011}. 

In this work, we perform local linear and nonlinear gyrokinetic simulations at two radial surfaces located in the tokamak core at $r/a=0.5$ (depicted as a red line in Fig.~\ref{fig:equilibrium}) and $r/a=0.6$. The safety factor and magnetic shear profiles in a region around $r/a=0.5$ are also shown in Fig.~\ref{fig:equilibrium}.
A Miller parameterisation~\cite{miller1998}, obtained by fitting the two radial surfaces using the \texttt{pyrokinetics} python library~\cite{pyrokinetics}, is considered in the following.
Local parameters at $r/a=0.5$ and $r/a=0.6$ are reported in table~\ref{tab:equilibrium}. In the following, we consider the surface at $r/a=0.5$, while the surface at $r/a=0.6$ is discussed in Sec.~\ref{sec:rad06}.

\begin{table}
    \centering
    \begin{tabularx}{0.6\textwidth}{ |>{\centering\arraybackslash}X|>{\centering\arraybackslash}X|>{\centering\arraybackslash}X|}
    \hline
    \multicolumn{3}{|c|}{\textbf{MAST \#22769}}\\
    \hline
    $r/a$       & 0.5 &   0.6 \\
    \hline
    $q$         & 1.07 &  1.20\\
    \hline
    $\hat{s}$   & 0.34 &  1.1\\
    \hline
    $\rho_*$    & 0.015 & 0.013\\
    \hline
    $\kappa$    & 1.41 & 1.42\\
    \hline
    $\delta$    & 0.23 & 0.14\\ 
    \hline
    $\Delta'$   &-0.13 & -0.17\\
    \hline
    $\beta_e$     & 0.057 & 0.049\\
    \hline
    $(a/c_s)\nu_e$ &   0.8 & 1.2\\
    \hline
    $n_e$ [$10^{19}$~m$^{-3}$] & 3.6 & 3.5\\
    \hline 
    $T_e$ [eV] & 450 & 370\\
    \hline
    $a/L_{n}$ & 0.22 & 0.31\\
    \hline
    $a/L_{T_e}$ & 2.1 & 2.2\\
    \hline
    $a/L_{T_D}$ & 1.7 & 2.0\\
    \hline
    \end{tabularx}
    \caption{Local parameters of the MAST discharge \#22769~\cite{valovivc2011} at the radial surfaces corresponding to $r/a = 0.5$ (reference surface) and $r/a = 0.6$ *this surface is discussed in Sec.~\ref{sec:rad06}). The parameters $\delta$, $\kappa$, $\Delta'$, $\nu_e$, $a/L_n$ and $a/L_T$ denote the plasma triangularity, the elongation, the Shafranov shift, the electron collision frequency, and the normalised inverse gradient lengths for density and temperature, respectively, with $\rho_*=\rho_s/a$.}
    \label{tab:equilibrium}
\end{table}

\begin{figure}
    \centering
    \subfloat[]{\includegraphics[width=0.48\textwidth]{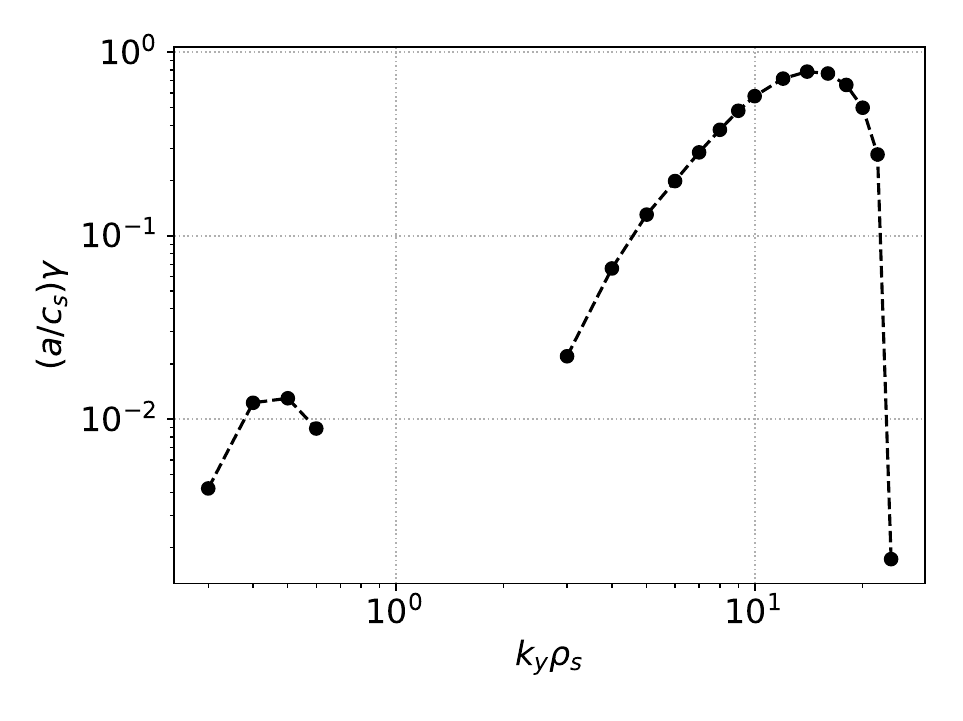}}\,
    \subfloat[]{\includegraphics[width=0.48\textwidth]{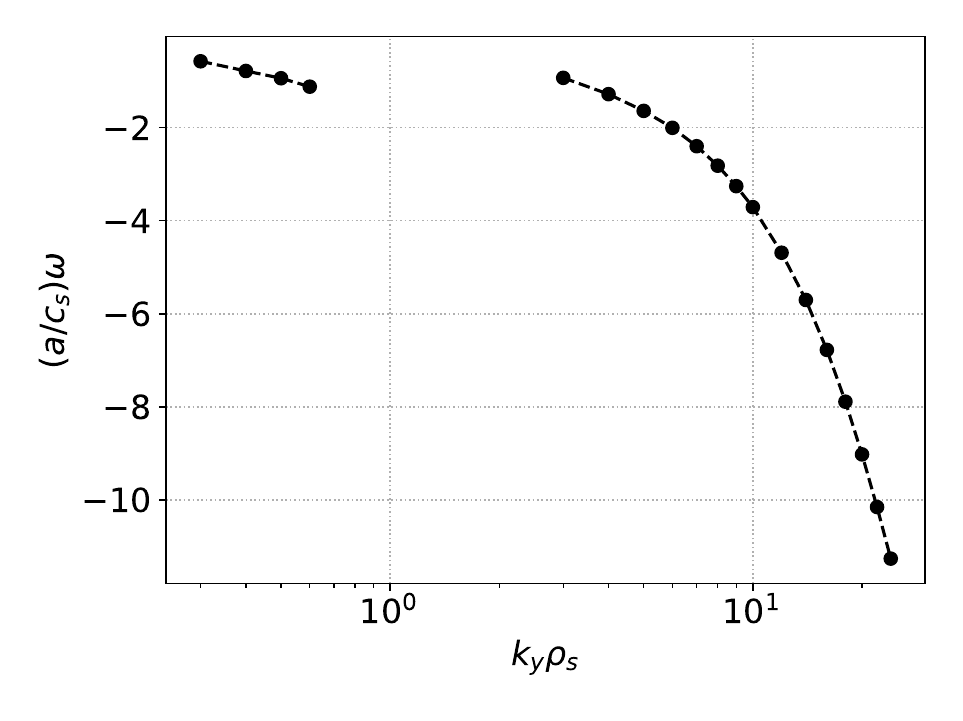}}
    \caption{Growth rate (a) and mode frequency (b) as a function of $k_y$ at $r/a=0.5$. Only unstable modes are shown. Results from linear GS2 simulations.}
    \label{fig:fullrange}
\end{figure}

Fig.~\ref{fig:fullrange} shows the growth rate and mode frequency of the dominant mode as a function of $k_y\rho_s$ from  linear simulations carried out by using the gyrokinetic code GS2~\cite{dorland2000,gs2}, with $k_y$ the binormal wave vector and $\rho_s=c_s/\Omega_{i}$ the ion sound Larmor radius. 
We note in Fig.~\ref{fig:fullrange} the presence of two different instabilities at ion and electron Larmor radius scale. The sign of the mode frequency of both instabilities is negative (negative sign is used here for mode phase velocity in the electron diamagnetic direction). The maximum growth rate of the electron scale instability is two orders of magnitude larger than the one of the ion scale instability.
The numerical resolution used in GS2 linear simulations is reported in table~\ref{tab:resolution}. Results of convergence tests are shown in \ref{sec:convergence}.

\begin{figure}
    \centering
    \subfloat[]{\includegraphics[width=0.45\textwidth]{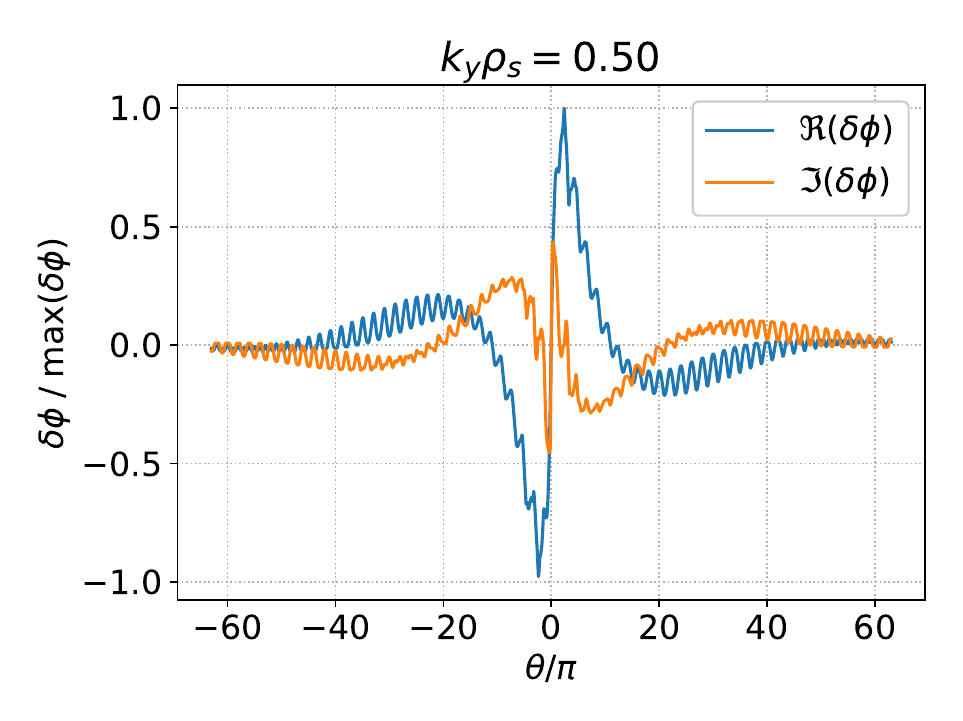}}\,
    \subfloat[]{\includegraphics[width=0.45\textwidth]{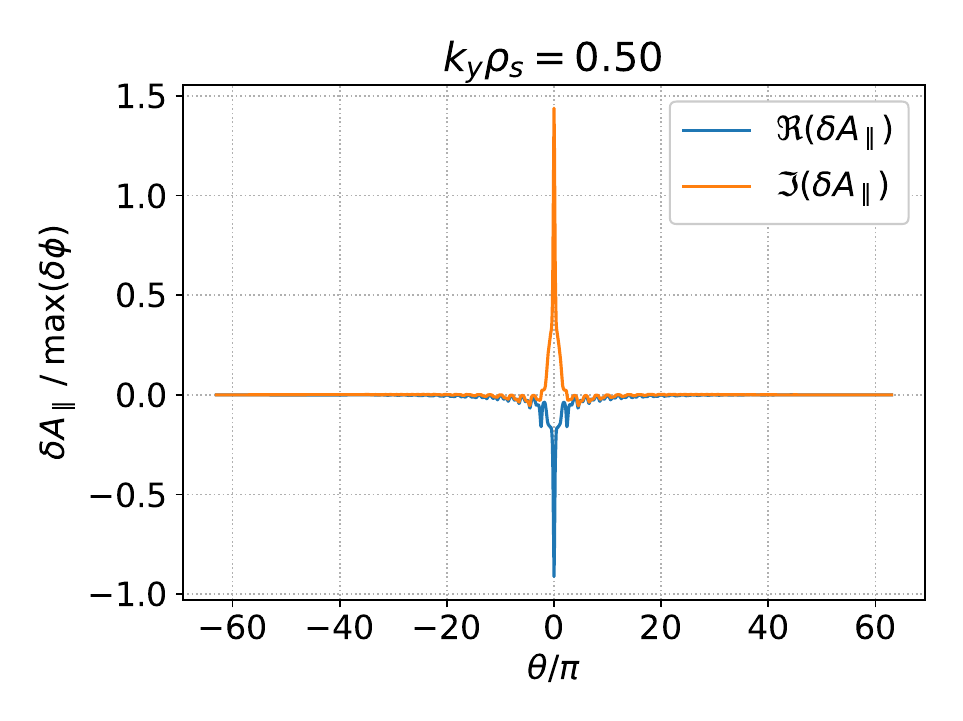}}\\
    \subfloat[]{\includegraphics[width=0.45\textwidth]{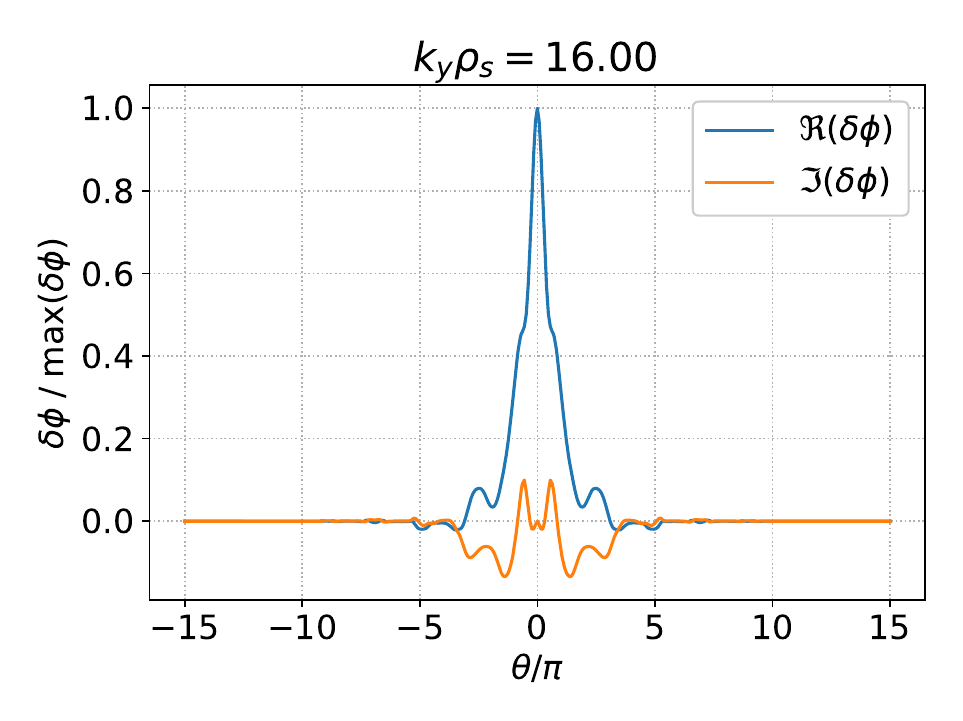}}\,
    \subfloat[]{\includegraphics[width=0.45\textwidth]{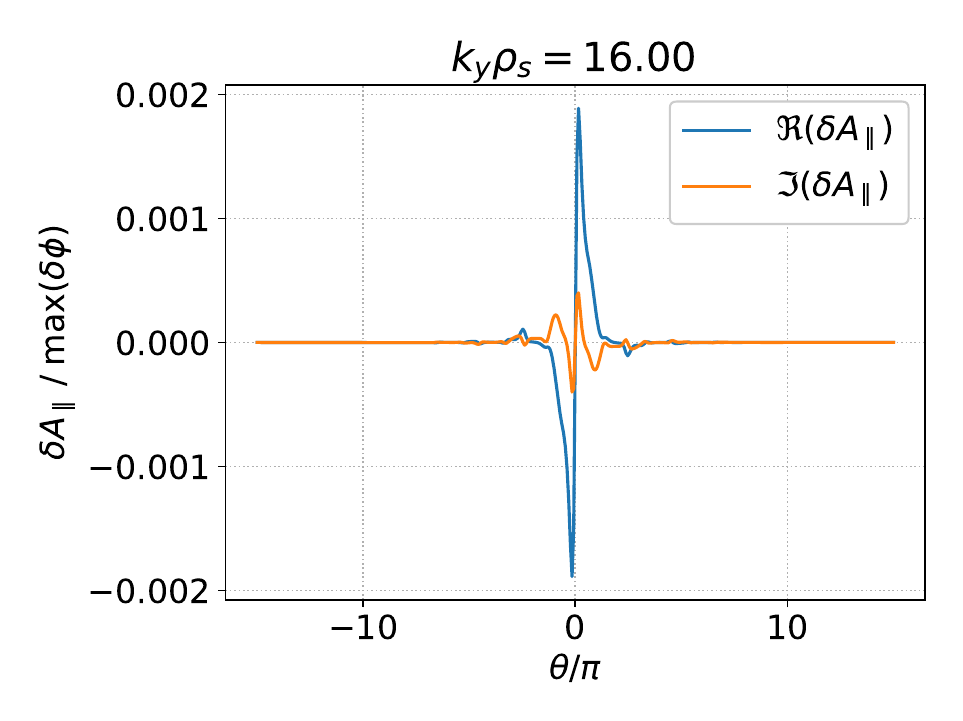}}
    \caption{Real and imaginary part of $\delta \phi/\max(\delta\phi)$ [(a) and (c)] and $\delta A_\parallel/\max(\delta\phi)$ [(b) and (d)] as a function of the ballooning angle at the two $k_y$ values corresponding to the maximum growth rate of the ion and electron scale instabilities. The fields $\delta\phi$ and $\delta A_\parallel$ are normalised to $\rho_*T_e/e$ and $\rho_*\rho_sB_0$, respectively.}
    \label{fig:eig_fullrange}
\end{figure}

The real and imaginary components of $\delta \phi$ and  $\delta A_\parallel$ are shown in Fig.~\ref{fig:eig_fullrange} as a function of the ballooning angle $\theta$ at the two different values of $k_y$ corresponding to the maximum growth rate of the ion and electron scale instabilities. 
The ion scale instability is characterised by $\delta \phi$ extended along $\theta$, while $\delta A_\parallel$ is very narrow around $\theta=0$. The mode has a tearing parity, i.e. $\delta \phi$ has odd parity and $\delta A_\parallel$ even parity with respect to $\theta = 0$. 
These are common features of MTMs~\cite{applegate2007}.
The mode at electron scale has a twisting parity ($\delta \phi$ even and $\delta A_\parallel$ odd) and both $\delta \phi$ and $\delta A_\parallel$ are localised in the region of $\theta = 0$. The electron scale modes are driven unstable by an ETG instability. This agrees with previous gyrokinetic linear simulations that have pointed out the presence of the ETG instability in various MAST scenarios~\cite{roach2009}.
We note that the amplitudes of $\delta \phi$ and $\delta A_\parallel$, both normalised to $\max(\delta\phi)$, are comparable at $k_y\rho_s=0.5$, while  $\delta\phi\gg\delta A_\parallel$ at $k_y\rho_s=20$. This is consistent with the ion (electron) scale instability being electromagnetic (electrostatic).

\begin{table}
    \centering
    \begin{tabular}{|c|c|c|c|}
    \hline
      &  \multicolumn{2}{c|}{\textbf{GS2}} & \textbf{CGYRO}\\
    \cline{2-4}
    & ETG & MTM & MTM\\
    \hline
    $n_\theta$  & 32 & 64 & 64\\
    \hline
    $n_r$     & 17 & 65  & 64\\
    \hline
    $n_\mathrm{\lambda}$, $n_\xi$   & 24  & 24  & 24\\
    \hline
    $n_\epsilon$          & 8 & 8 & 8\\
    \hline
    \end{tabular}
    \caption{Numerical resolution used in GS2 and CGYRO linear simulations, with $n_\theta$ and $n_r$ the number of grid points in the parallel and radial directions, respectively, and $n_\epsilon$ the number of the energy grid points. In GS2  $n_\mathrm{\lambda}$ is the number of pitch-angles, while in CGYRO $n_\xi$ is the number of Legendre pseudospectral meshpoints in the pitch-angle space. Results of CGYRO linear simulations are presented in Sec.~\ref{sec:linear}.}
    \label{tab:resolution}
\end{table}

In this work, we focus only on the MTM instability, even though, as demonstrated in \ref{sec:etg}, ETG modes drive most of the turbulent transport at $r/a=0.5$ in this particular MAST equilibrium. We highlight that the aim of this paper is mainly to investigate the saturation mechanism of this MTM instability and the role of the stochastic layer in MAST rather than to provide an accurate prediction of the heat flux in this MAST case. 

Since $\nu_e \simeq \omega$, this MTM instability sits between the collisionless and semi-collisional regimes of Ref.~\cite{drake1977}. This regime has been numerically studied in Ref.~\cite{gladd1980} and analytically addressed in a recent work reported in Ref.~\cite{zocco2015}.
Both works show that the growth rate of MTMs strongly depends on the electron collision frequency.
Since the collisionality in the MAST reference case is $\nu_*\simeq 0.1$,  trapped electron effects may also provide an additional drive for the MTM instability, as described in Ref.~\cite{catto1981}. 

\begin{figure}
    \centering
    \subfloat[]{\includegraphics[width=0.32\textwidth]{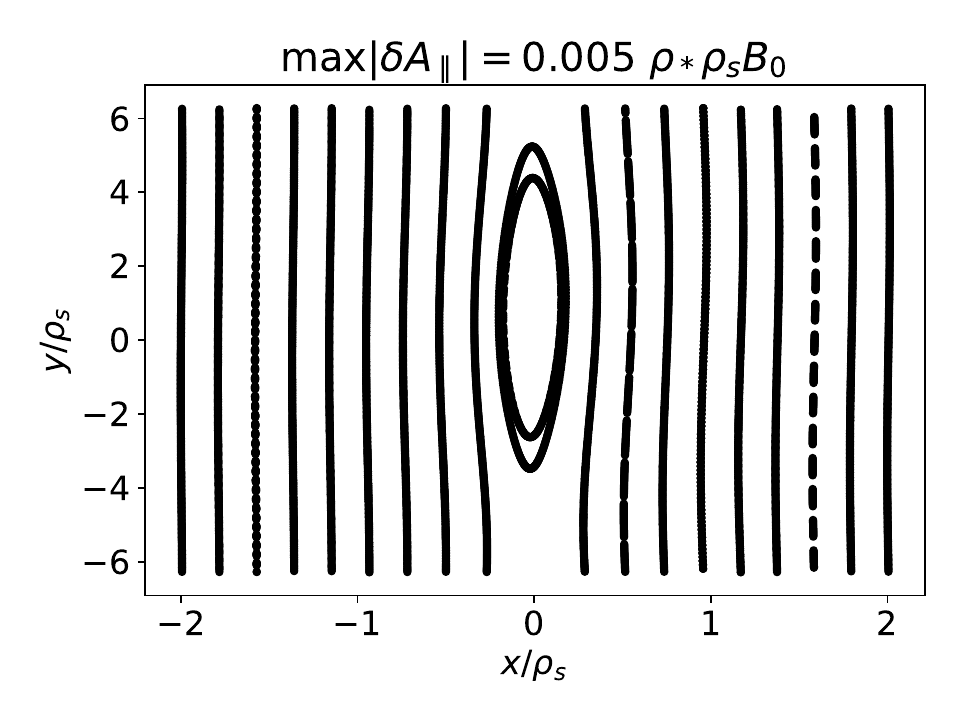}}\,
    \subfloat[]{\includegraphics[width=0.32\textwidth]{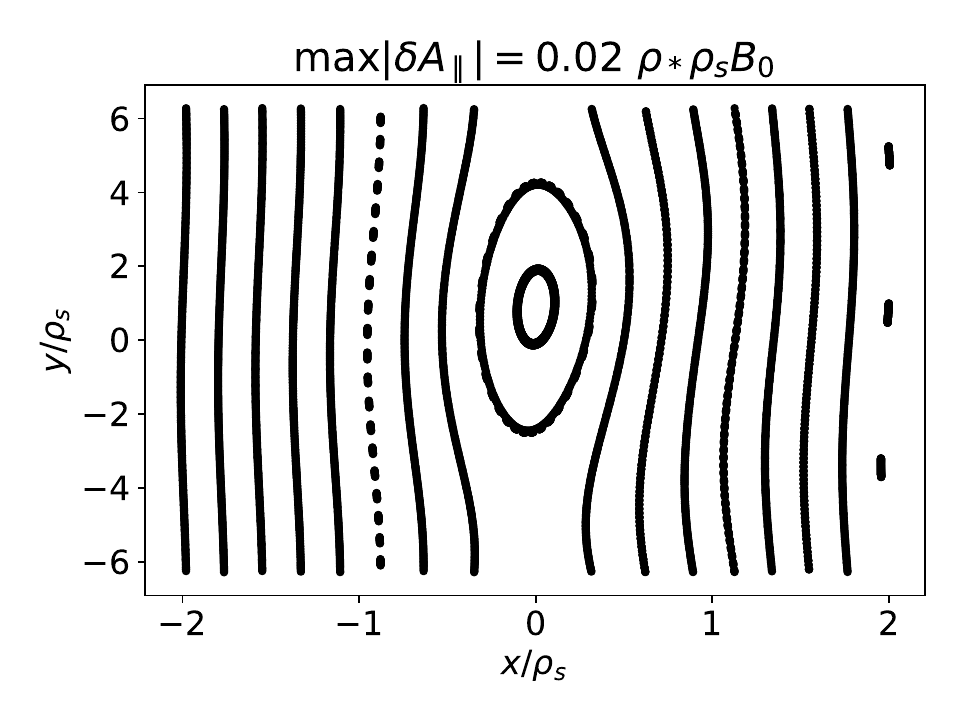}}\,
    \subfloat[]{\includegraphics[width=0.32\textwidth]{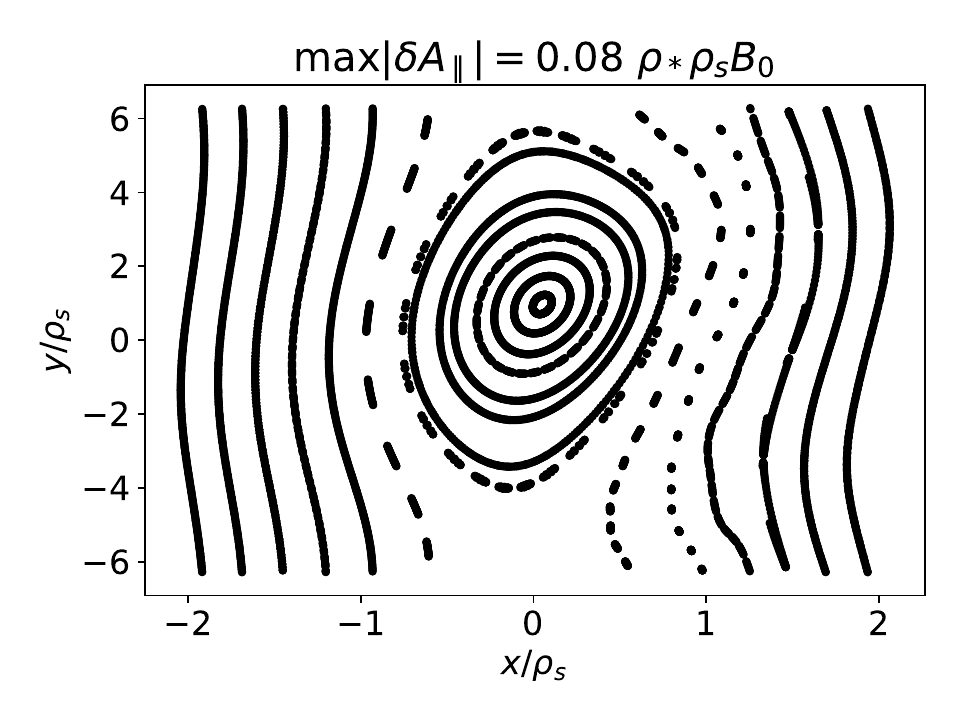}}
    \caption{Poincare map of the magnetic field lines in the proximity of a rational surface (centered at $x=0$ here) showing the formation of a magnetic island caused by the MTM instability at $k_y\rho_s = 0.5$. Panels (a), (b) and (c) correspond to different amplitudes of $\delta A_\parallel$.}
    \label{fig:island_linear}
\end{figure}

The MTM is a tearing instability that leads to the formation of magnetic islands.  Fig.~\ref{fig:island_linear} shows a Poincaré map of the magnetic field at three different amplitudes of $\delta A_\parallel$ at $k_y\rho_s=0.5$, which corresponds to the most unstable MTM.
We highlight that the amplitude of $\delta A_\parallel$ used in Fig.~\ref{fig:island_linear} is chosen only for representative purposes, as the actual value of $\delta A_\parallel$ for a linear unstable mode grows exponentially until nonlinear effects cause saturation. 
The magnetic island forms at a rational surface located at $x=0$. We note from Fig.~\ref{fig:island_linear} that the magnetic island width increases from $w_\mathrm{island} \simeq 0.5\,\rho_s$ to $w_\mathrm{island} \simeq 2\,\rho_s$ as $\delta A_\parallel$ is increased from $5\times 10^{-3}\,\rho_*\rho_sB_0$ to $8\times 10^{-2}\,\rho_*\rho_sB_0$, which is in agreement with the prediction of Eq.~(\ref{eqn:wisland}).
The distance between adjacent rational surfaces at $k_y\rho_s = 0.5$ is given by Eq.~(\ref{eqn:sep_single}) and it is $\Delta r = 1/(sk_y)\simeq 6\,\rho_s$, which is a factor of three larger than the island width at $\max|\delta A_\parallel| = 0.08\,\rho_*\rho_sB_0$. If $\delta A_\parallel$ is further increased, the magnetic islands generated by this mode at different rational surfaces overlap partially, therefore generating a region of stochastic magnetic field. 
The formation of a stochastic layer can strongly enhance heat transport, as shown later in Sec.~\ref{sec:islands}. We note that multiple toroidal modes are evolved in nonlinear simulations, thus reducing the distance between adjacent resonant surfaces as compared to the distance between rational surfaces for fixed $n$.

\section{Linear characterisation of the MTM instability}
\label{sec:linear}

\begin{figure}
    \centering
    \subfloat[]{\includegraphics[width=0.48\textwidth]{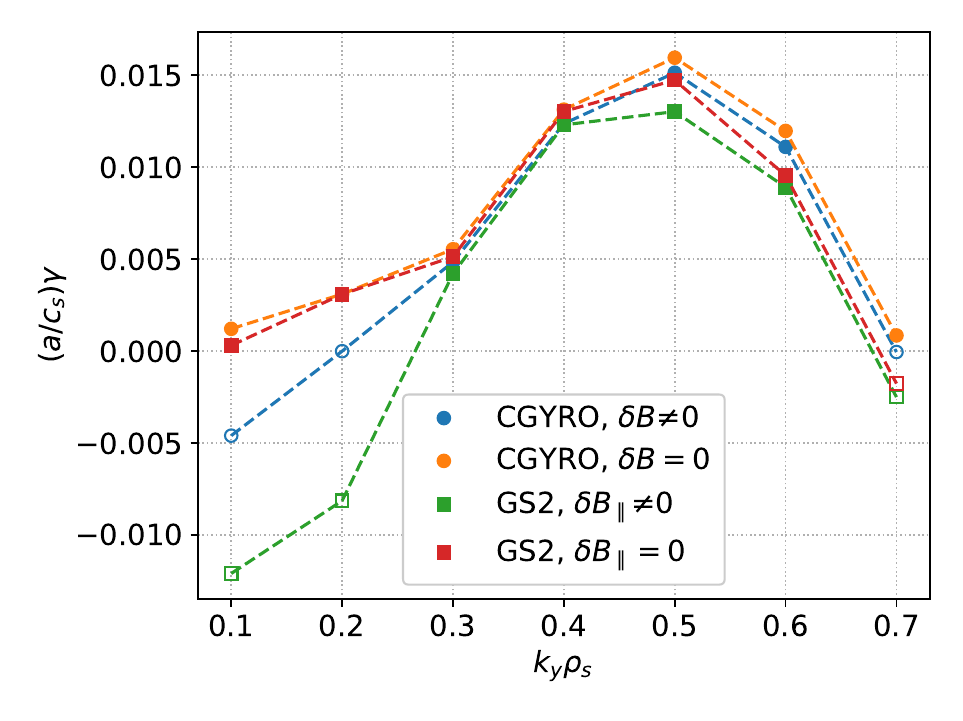}}\,
    \subfloat[]{\includegraphics[width=0.48\textwidth]{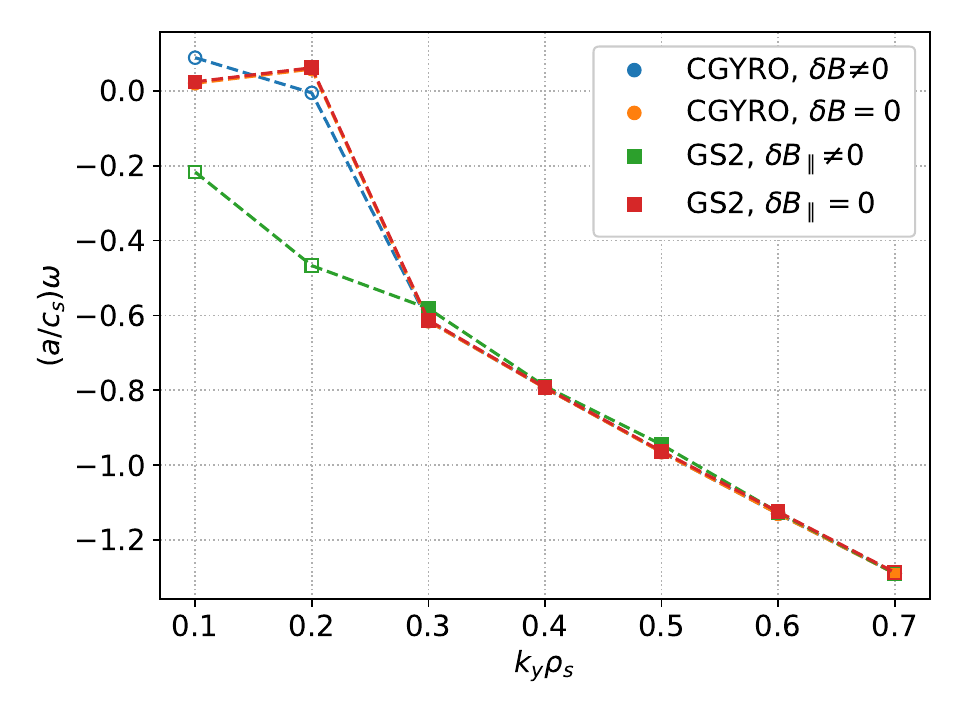}}
    \caption{Growth rate (a) and mode frequency (b) as a function of $k_y$ from CGYRO and GS2 linear simulations with and without $\delta B_\parallel$ at $r/a=0.5$. Unstable and stable modes are shown with solid and open markers, respectively. }
    \label{fig:ref}
\end{figure}

We explore here the sensitivity of the MTM instability to various parameters and, in particular, to the electron collision frequency. 
Linear simulations are carried out with the gyrokinetic codes CGYRO~\cite{candy2016} and GS2, using a similar numerical resolution in the two codes, as reported in table~\ref{tab:resolution}. 
A benchmark of the reference case is shown in Fig.~\ref{fig:ref}, where linear simulations with and without $\delta B_\parallel$ are considered.
A good agreement between CGYRO and GS2 growth rate and mode frequency values is observed in the region $0.2<k_y\rho_s<0.7$ where MTMs are unstable. 
We note that the MTM instability is weakly affected by $\delta B_\parallel$, in agreement with Ref.~\cite{applegate2007}. 
The modes at $k_y\rho_s\le 0.2$ are stable when $\delta B_\parallel \ne 0$ and unstable when $\delta B_\parallel = 0$.
Parallel magnetic fluctuations are therefore important to suppress this low $k_y$ ion temperature gradient (ITG) instability, which has a positive mode frequency sign (phase velocity in the ion diamagnetic direction) and is also observed in the electrostatic limit.
Parallel magnetic fluctuations are retained in all the following linear and nonlinear simulations, and therefore the ITG mode at low $k_y$ is stable.

\begin{figure}
    \centering
    \subfloat[]{\includegraphics[width=0.48\textwidth]{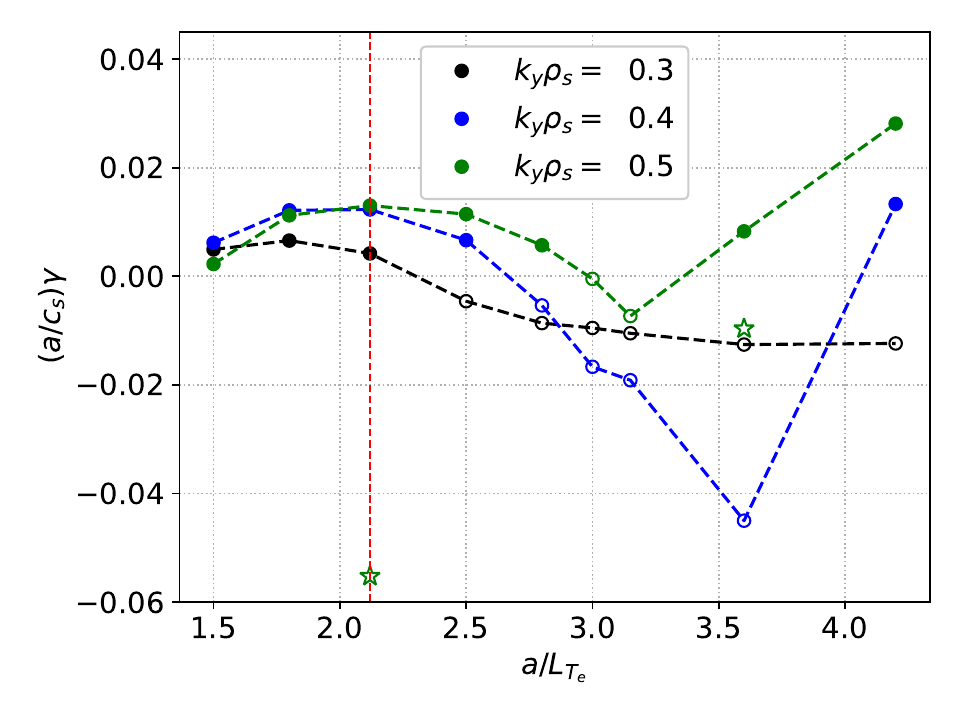}}\,
    \subfloat[]{\includegraphics[width=0.48\textwidth]{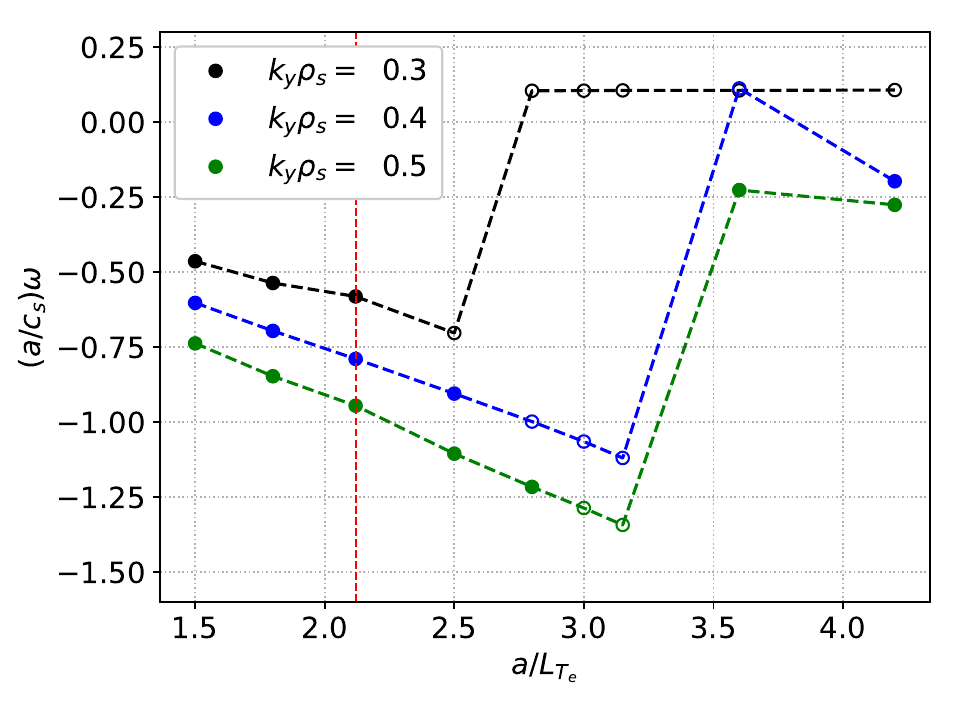}}
    \caption{Growth rate (a) and mode frequency (b) as a function of the temperature gradient at three different values of $k_y$ at $r/a=0.5$. Unstable and stable modes are shown with solid and open markers, respectively. The star markers represent simulations with adiabatic passing and kinetic trapped electrons at $k_y\rho_s=0.5$ with $a/L_{T_e} = 2.1$ (nominal value) and $a/L_{T_e} = 3.6$. The red vertical dashed line corresponds to the reference value of $a/L_{T_e}$. Results from GS2 linear simulations.}
    \label{fig:lte}
\end{figure}

The effect of the electron temperature gradient is investigated in Fig.~\ref{fig:lte}, where growth rate and mode frequency values are shown at different values of $a/L_{T_e}$. The growth rate depends on the electron temperature gradient, as predicted by early analytical works~\cite{hazeltine1975,hassam1980}.
We note that the dependence on $a/L_{T_e}$ is non-monotonic and the local maximum growth rate is reached at the reference value of $a/L_{T_e}$. A non-monotonic dependence has also been observed in previous MAST linear simulations, as shown in Ref.~\cite{applegate2007}, where a resonance mechanism occurring at $\nu_e \simeq \omega$ is proposed as a possible explanation, as well as in ASDEX Upgrade and JET linear gyrokinetic simulations~\cite{moradi2013}.
In addition, we also show in Fig.~\ref{fig:lte} the growth rate value from linear simulations with adiabatic passing electrons and kinetic trapped electrons at two different values of $a/L_{T_e}$. We note that these modes are stable when adiabatic passing electrons are considered, hence pointing out a minor role played by trapped electrons.

\begin{figure}
    \centering
    \subfloat[]{\includegraphics[width=0.48\textwidth]{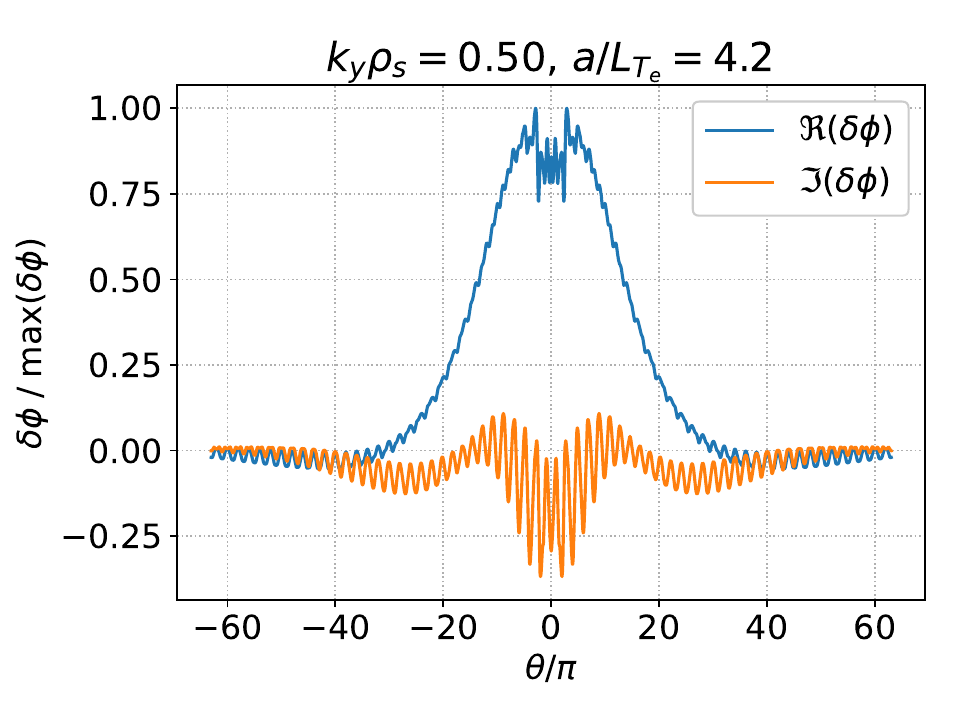}}\,
    \subfloat[]{\includegraphics[width=0.48\textwidth]{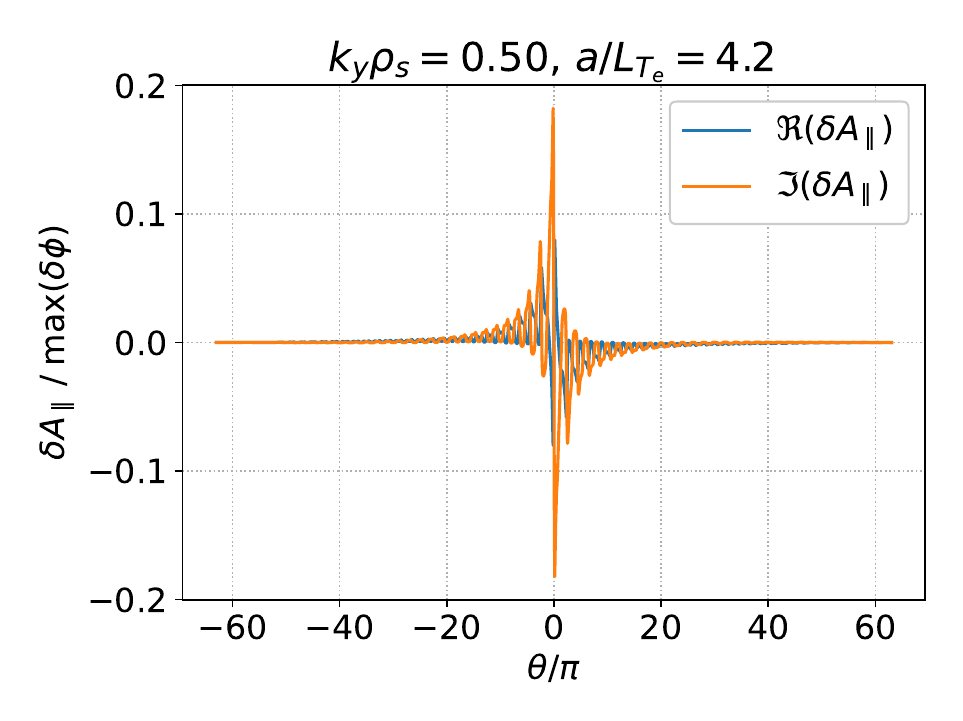}}
    \caption{Real and imaginary part of $\delta \phi/\max(\delta \phi)$ (a) and $\delta A_\parallel/\max(\delta \phi)$ (b) corresponding to the unstable mode at $k_y\rho_s=0.5$ and  $a/L_{T_e} = 4.2$. The fields $\delta\phi$ and $\delta A_\parallel$ are normalised to $\rho_*T_e/e$ and $\rho_*\rho_sB_0$, respectively.}
    \label{fig:eig_etg}
\end{figure}

Fig.~\ref{fig:lte} shows that MTMs are stable at intermediate $a/L_{T_e}$ values, while another instability appears at large $a/L_{T_e}$, associated with a transition in the mode frequency, which remains in the electron diamagnetic direction.  
The eigenfunctions corresponding to the mode at $k_y\rho_s = 0.5$ and $a/L_{T_e} = 4.2$ are shown in Fig.~\ref{fig:eig_etg}. The electrostatic potential is very elongated in the ballooning angle, similarly to the MTM instability. On the other hand, the mode has a twisting parity ($\delta \phi$ is even and $\delta A_\parallel$ odd), it is unstable also in the electrostatic limit and it is driven unstable by kinetic passing electrons, as shown in Fig.~\ref{fig:lte}.
The instability appearing at large $a/L_{T_e}$ is an ETG mode characterised by $k_y\rho_s\sim 1$ and $k_x\rho_s>1$, which is similar to the long wavelength ETG instability described in Ref.~\cite{hardman2022} (see \ref{sec:etg_linear} for further details on this instability).    

\begin{figure}
    \centering
    \subfloat[]{\includegraphics[width=0.48\textwidth]{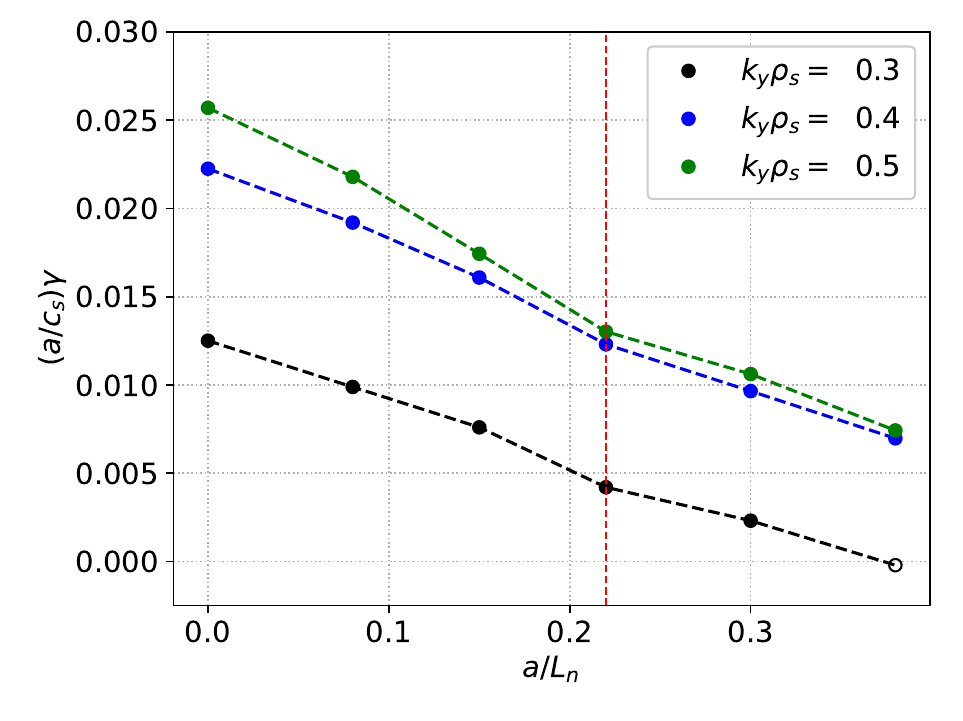}}\,
    \subfloat[]{\includegraphics[width=0.48\textwidth]{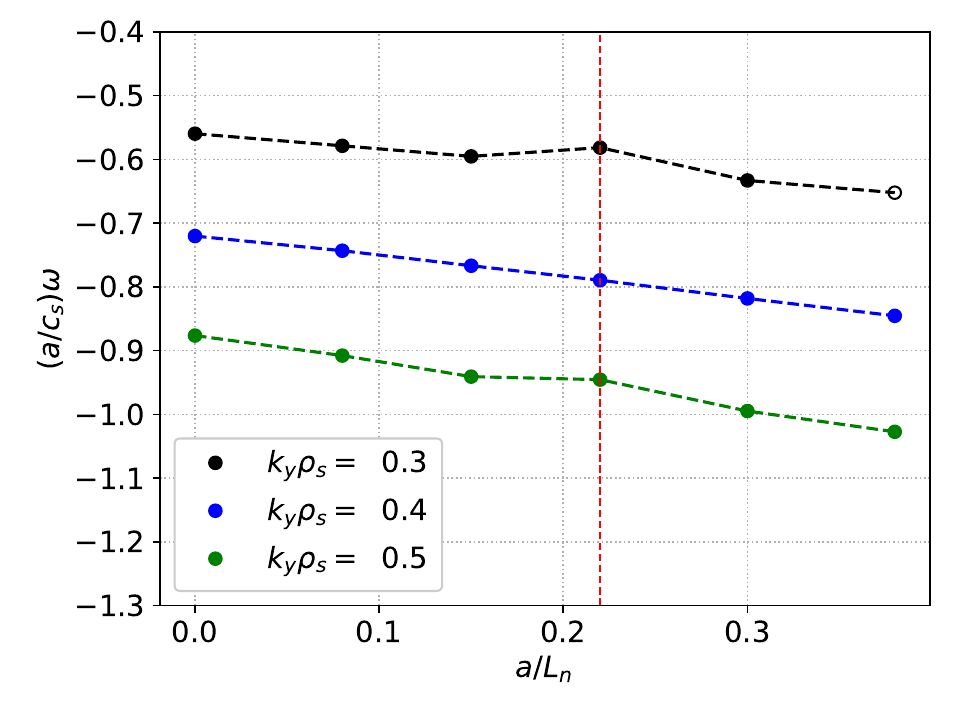}}
    \caption{Growth rate (a) and mode frequency (b) as a function of the density gradient at three different values of $k_y$ at $r/a=0.5$. Unstable and stable modes are shown with solid and open markers, respectively. The red vertical dashed line corresponds to the reference value of $a/L_n$. Results from GS2 linear simulations.}
    \label{fig:ln}
\end{figure}

The results of a density gradient scan are presented in Fig.~\ref{fig:ln}, where the growth rate and mode frequency at different $k_y$ values are shown as a function of $a/L_n$.
The growth rate decreases as the density gradient increases, similar to recent findings in linear simulations of MTMs for a high-$\beta$ spherical tokamak conceptual power plant design~\cite{patel2021}. 

\begin{figure}
    \centering
    \subfloat[]{\includegraphics[width=0.48\textwidth]{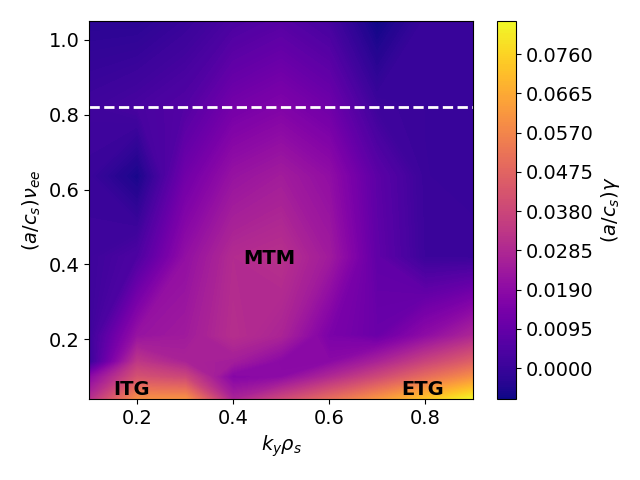}}\,
    \subfloat[]{\includegraphics[width=0.48\textwidth]{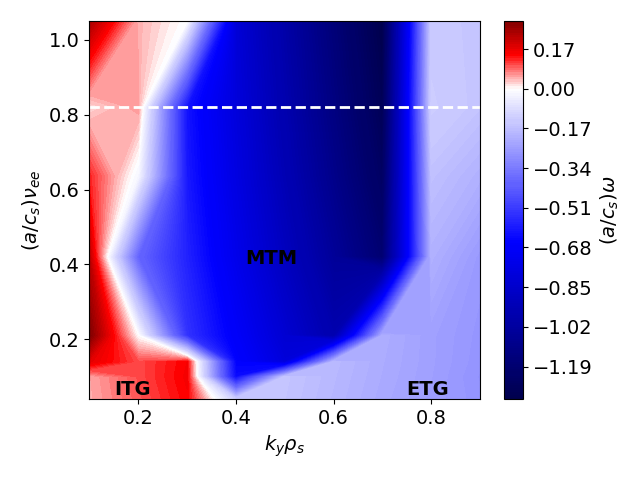}}
    \caption{Growth rate (a) and mode frequency (b) as a function of $k_y$ and $\nu_e$ at $r/a=0.5$. The white dashed horizontal line indicates the value of $\nu_e$ in the reference MAST case. Results from CGYRO linear simulations.}
    \label{fig:coll2D}
\end{figure}

\begin{figure}
    \centering
    \subfloat[]{\includegraphics[width=0.48\textwidth]{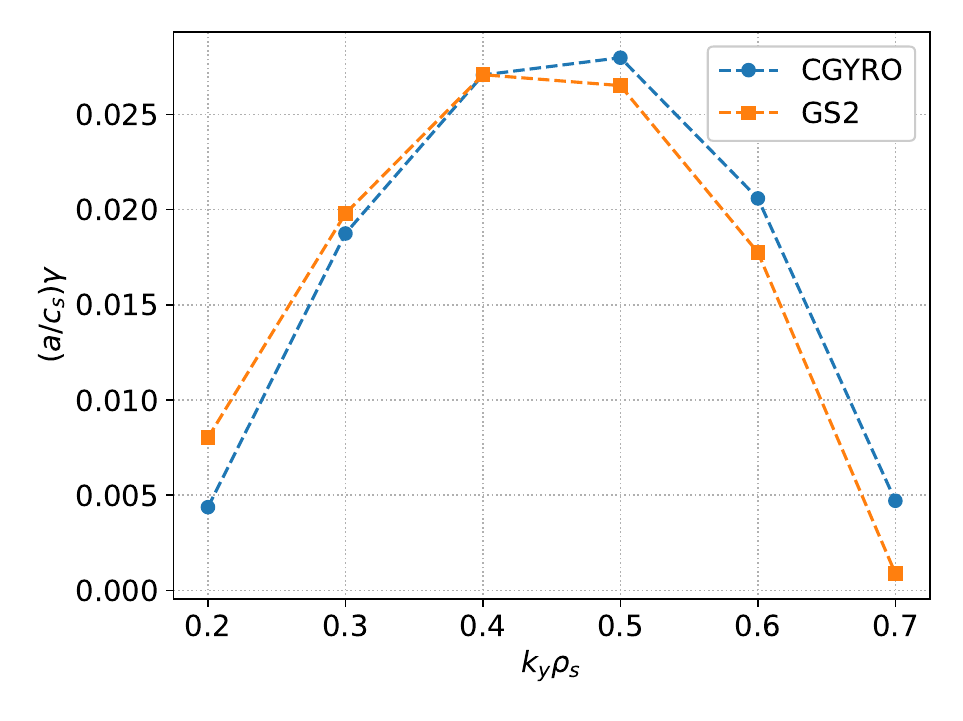}}\,
    \subfloat[]{\includegraphics[width=0.48\textwidth]{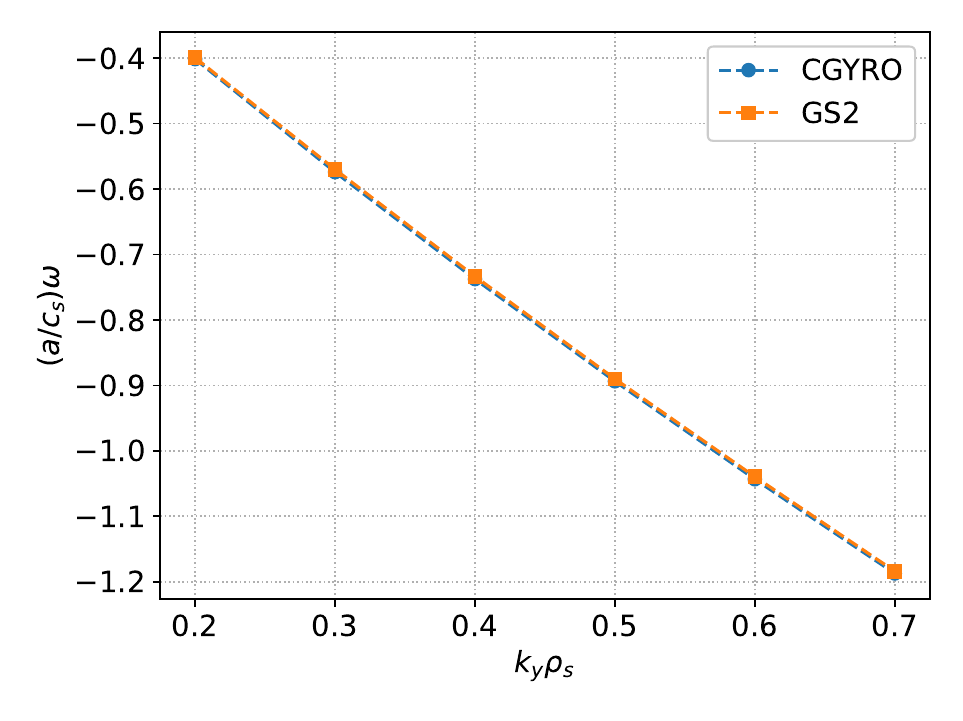}}
    \caption{Comparison between CGYRO and GS2 growth rate (a) and mode frequency (b) from linear simulations with $\nu_e = 0.42\,c_s/a$ at $r/a=0.5$. Only unstable modes are shown.}
    \label{fig:coll1D}
\end{figure}

The effect of the electron collision frequency is investigated in Fig.~\ref{fig:coll2D}, where growth rate and frequency values are shown at various values of $\nu_e$ and $k_y$. The MTM instability is suppressed at low collisionality in favour of ITG (at low $k_y$) and ETG (at high $k_y$), thus confirming the collisional nature of this MTM instability, which is stable in the collisionless limit. The MTM growth rate increases with $\nu_e$, until it reaches its maximum value around $\nu_e = 0.42\ c_s/a$. At this value of collision frequency, which is approximately half the value of the collision frequency in the reference case, the growth rate is a factor of two larger than in the reference case. The growth rate value decreases when the collisionality is further increased. Since the maximum growth rate occurs at $\nu_e \simeq \omega_{*e}$, with $\omega_{*e}$ the diamagnetic electron frequency, this may suggest a resonance mechanism similar to the one observed in the electron temperature scan. We note that the ITG instability is suppressed when $\nu_e > 0.4\ c_s/a$, while the onset of the ETG instability shifts at higher $k_y$ values when $\nu_e$ is increased (see \ref{sec:etg_linear} for further details). 
We also note that future magnetic confinement fusion devices, based on the high-$\beta$ spherical tokamak concept, are expected to have a lower collisionality than the case considered here and, therefore, the corresponding scenario may be located in the phase space region where the collisional MTM growth rate increases with the electron collision frequency and where the MTM instability connects to other instabilities, such as ITG or ETG modes, although the higher $\beta$ values expected in high-$\beta$ reactor-scale spherical tokamak scenarios might change the location and/or the presence of a maximum of $\gamma(\nu_e)$.   
In the following, we consider the most unstable case at $r/a=0.5$ occurring when $\nu_e = 0.42\ c_s/a$. A comparison between CGYRO and GS2 linear simulations at $\nu_e = 0.42\ c_s/a$ is shown in Fig.~\ref{fig:coll1D}. A good agreement is observed over the entire $k_y$ range both in the growth rate and mode frequency values.

\begin{figure}
    \centering
    \subfloat[]{\includegraphics[width=0.48\textwidth]{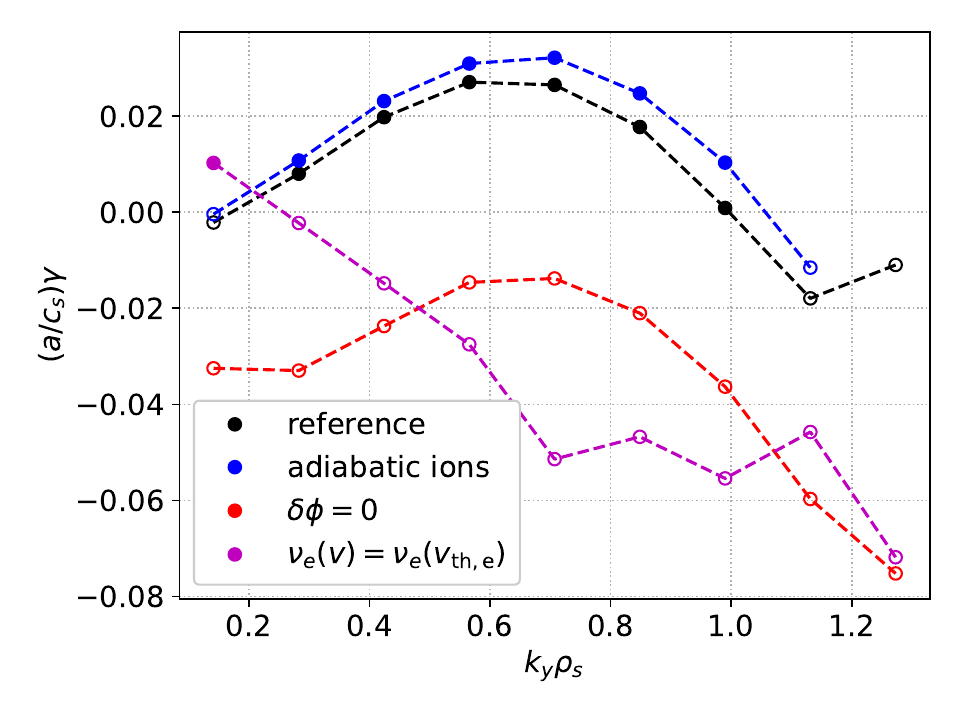}}\,
    \subfloat[]{\includegraphics[width=0.48\textwidth]{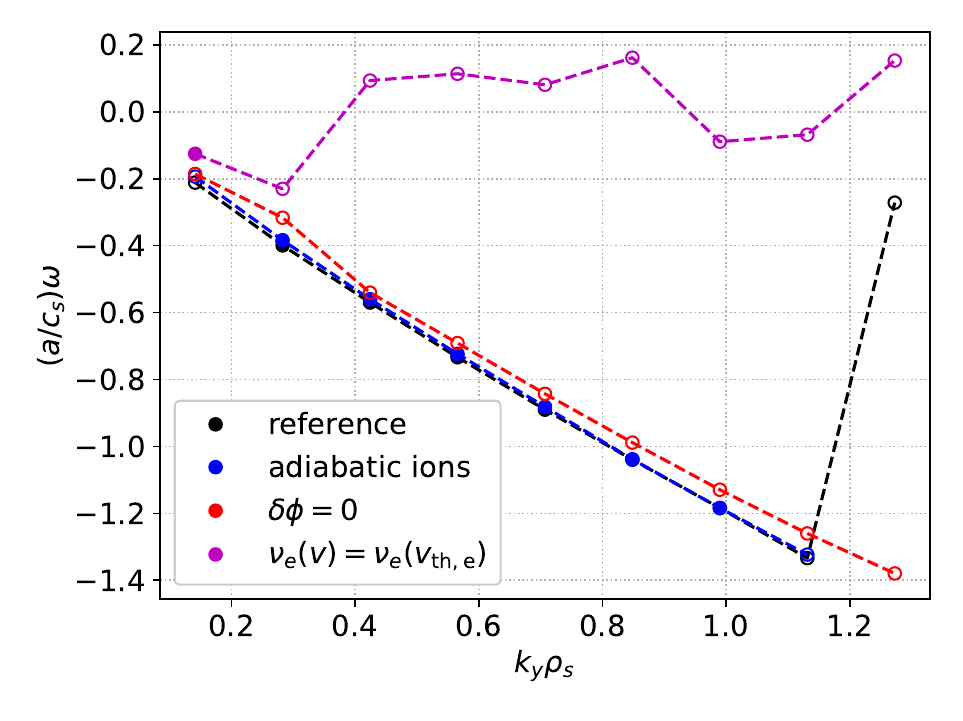}}
    \caption{Growth rate (a) and mode frequency (b) as a function of $k_y$ from GS2 linear simulations at $\nu_e = 0.42\,c_s/a$ with adiabatic ions (blue line), with $\delta\phi = 0$ (red line) and with a velocity independent collision frequency (magenta line). Solid and open markers are used for unstable and stable modes, respectively.}
    \label{fig:test_mtm}
\end{figure}

The driving mechanism of the MTM instability of Ref.~\cite{hazeltine1975} requires a velocity dependent collision frequency~\cite{hassam1980}. In Fig.~\ref{fig:test_mtm}, we compare the results of linear simulations with and without a velocity dependence in the collisional operator. The MTM instability is retrieved only when the velocity dependence is retained. Modes at $k_y\rho_s>0.1$ are stable when $\nu_e(v) = \nu_e(v_\mathrm{th, e})$. The mode at $k_y\rho_s=0.1$ is stable in the reference case and unstable when an energy independent collision operator is considered. 
As an aside, we note, however, that earlier linear gyrokinetic studies of MAST in Ref.~\cite{applegate2007} found that MTMs remained unstable when the energy dependence of the collision operator was similarly removed.

Previous analytical works suggest also an important stabilising effect on the MTM instability from the ion dynamics~\cite{cowley1986}. Results of linear simulations with adiabatic ions are shown in Fig.~\ref{fig:test_mtm}. 
Growth rate values increase slightly when considering adiabatic ions, so the stabilising effect from kinetic ions is very weak in the case considered here.

Finally, the dependence on the inclusion of the electrostatic potential fluctuations is investigated. Fig.~\ref{fig:test_mtm} shows that MTMs are stable in the simulation with $\delta \phi = 0$, i.e. electrostatic potential fluctuations are essential for the drift-tearing mode to be unstable in this case. This is in agreement with the numerical calculations of Ref.~\cite{gladd1980}. We note that inclusion of $\delta \phi$ was also found to be destabilising in previous MAST MTMs linear simulations~\cite{applegate2007}. This is expected from theoretical calculations in Refs.~\cite{drake1977,zocco2015}, where in Ref.~\cite{drake1977} it is noted that the electrostatic potential plays an increasing role for MTMs in more collisional limits.

\section{Nonlinear simulation results and saturation mechanism}
\label{sec:nonlinear}

We present here the results of the first nonlinear simulations of MTM turbulence carried out in a MAST case. The simulations are performed using CGYRO at the radial surface located at $r/a=0.5$.
Nonlinear simulations are carried out without equilibrium $\mathbf{E}\times\mathbf{B}$ flow shear, whose effect is investigated in Ref.~\cite{patel2023a}, which shows that it has a relatively weak effect on the saturated heat flux value in this MAST reference case.
We note that these nonlinear simulations are computationally quite expensive since a high numerical resolution is required to properly resolve the MTM instability, especially in the radial direction (see table~\ref{tab:resolution_nl}). The simulation with the highest numerical resolution considered in this work required approximately $10^5$ CPU-hours on the ARCHER2 supercomputer (Edinburgh, United Kingdom). 

\begin{table}
    \centering
    \begin{tabularx}{0.8\textwidth}{ |>{\centering\arraybackslash}X|>{\centering\arraybackslash}X|>{\centering\arraybackslash}X|>{\centering\arraybackslash}X|}
    \hline
     \multicolumn{4}{|c|}{\textbf{CGYRO nonlinear simulations}}\\
     \hline
     Parameters & Reference & Lower $k_{y, \mathrm{min}}$ & Higher $\hat{s}$ \\
    \hline
    $n_\theta$  & 32 & 32 & 32 \\
    \hline
    $n_r$     &  256 &  256 &  256\\
    \hline
    $n_{k_y}$ & 10 & 20 & 16\\
    \hline
    $n_\xi$   & 24 &  24 &  24\\
    \hline
    $n_\epsilon$ & 8 & 8 & 8\\
    \hline
    $k_{y,\mathrm{min}}\rho_s$ & 0.07 & 0.035 & 0.035\\
    \hline
    $L_x/\rho_s$ & 168 & 168 & 82 \\
    \hline
    \end{tabularx}
    \caption{Numerical resolution used in CGYRO nonlinear simulations. The quantities $n_k{_y}$, $k_{y,\mathrm{min}}$ and $L_x$ represent the number of evolved $k_y$ modes, the minimum evolved finite $k_y$ value and the radial extent of the flux tube domain, respectively. The simulation ``higher $\hat{s}$'' is discussed in Sec.~\ref{sec:islands}.}
    \label{tab:resolution_nl}
\end{table}

\begin{figure}
    \centering
    \subfloat[]{\includegraphics[width=0.48\textwidth]{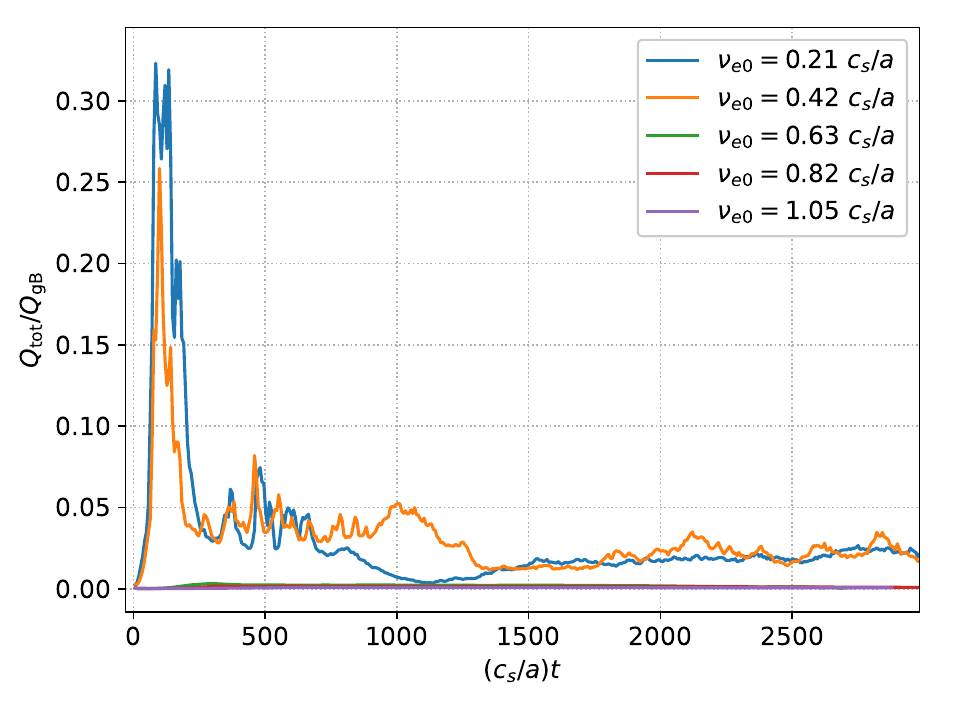}}\,
    \subfloat[]{\includegraphics[width=0.48\textwidth]{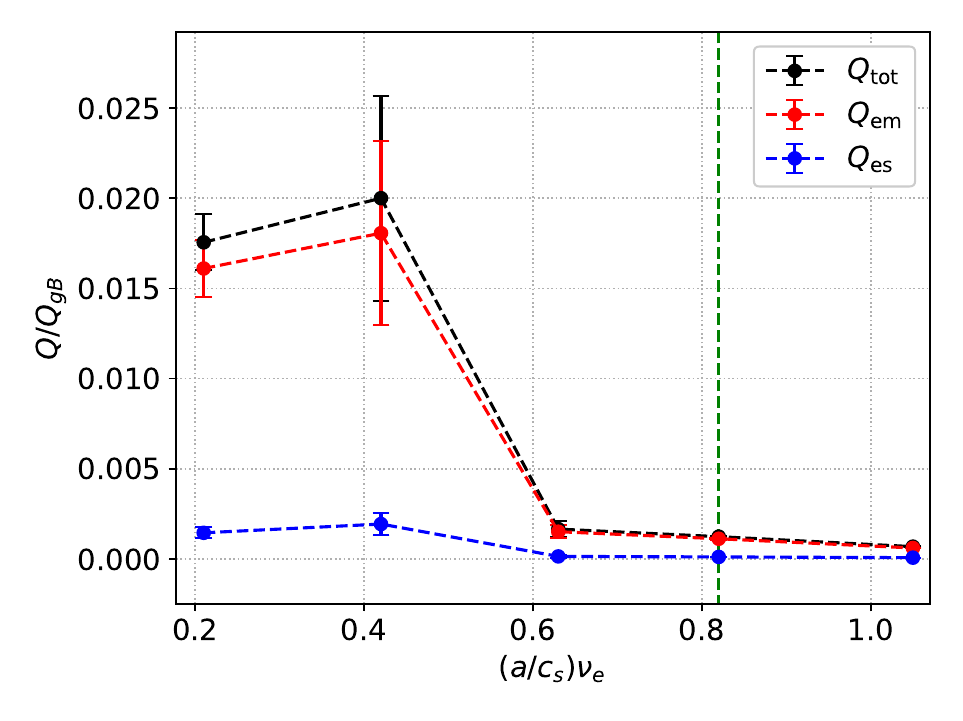}}
    \caption{(a) Time trace of total heat flux from CGYRO nonlinear simulations with different values of electron collision frequency at $r/a=0.5$. The nominal value of collision frequency in the  MAST reference case is $\nu_e = 0.82\ c_s/a$. (b) Saturated level of the total (black), electromagnetic (red) and electrostatic (blue) heat flux obtained by time averaging the heat flux at $t>1500\,a/c_s$. The error bars are determined from standard deviation. The green vertical dashed line denotes the experimental value of $\nu_e$ at $r/a=0.5$.}
    \label{fig:q_nu}
\end{figure}

Fig.~\ref{fig:q_nu}~(a) shows the time trace of the total heat flux from nonlinear simulations with different values of $a\nu_e/c_s\in \{1.05, 0.82, 0.63, 0.42, 0.21\}$. The value of the electron collision frequency in the MAST reference case at $r/a=0.5$ is $\nu_e = 0.82\ c_s/a$. The linear scan presented in the previous section shows that the maximum MTM growth rate is achieved at approximately $\nu_e \simeq 0.42\ c_s/a$.
We note that the heat flux driven by MTMs is negligible for $\nu_e > 0.6\ c_s/a$, despite MTMs being linearly unstable (see Fig.~\ref{fig:coll2D}).
Therefore, there is no contribution to the heat flux from the MTM instability in the reference MAST case at $r/a=0.5$ (most of the turbulent heat flux is driven by the ETG instability as shown in \ref{sec:etg}) despite MTMs being unstable. This shows that the presence of linearly unstable MTMs is not a sufficient condition to drive significant heat flux. 
Fig.~\ref{fig:q_nu}~(b) shows the electromagnetic and electrostatic contribution to the saturated heat flux at different values of electron collision frequency.  When $\nu_e$ decreases from $0.63\ c_s/a$ to $0.42\ c_s/a$, the heat flux increases by an order of magnitude, while the maximum growth rate of the linear MTM instability increases by less than a factor two. 
At $\nu_e = 0.42\ c_s/a$ and $\nu_e = 0.21\ c_s/a$, the heat flux saturates approximately at $Q_\mathrm{tot}\simeq 0.02\ Q_{gB}$, where $Q_{gB} = \rho_*^2 n_e T_e c_s$ is the gyro-Bohm heat flux, corresponding to $Q_\mathrm{tot}\simeq 0.002$~MW/m$^2$. 
We note that the heat flux at $\nu_e < 0.6\ c_s/a$ saturates at a value that is more than a factor of two smaller than the saturated heat flux driven by the ETG instability in the reference case (see \ref{sec:etg}). 
Fig.~\ref{fig:q_nu}~(b) shows that the electromagnetic heat flux largely dominates over the electrostatic contribution, in agreement with previous nonlinear gyrokinetic simulations of MTMs~\cite{guttenfelder2011,doerk2011}. 

Since the aim of the present work is to analyse the properties of MTM turbulence in cases built from a MAST scenario, from this point onwards we focus  our studies on more strongly driven MTM turbulence at $r/a=0.5$ that arises at $\nu_e=0.42\,c_s/a$ (which is half the nominal collisionality on this surface).

\begin{figure}
    \centering
    \includegraphics[scale=0.6]{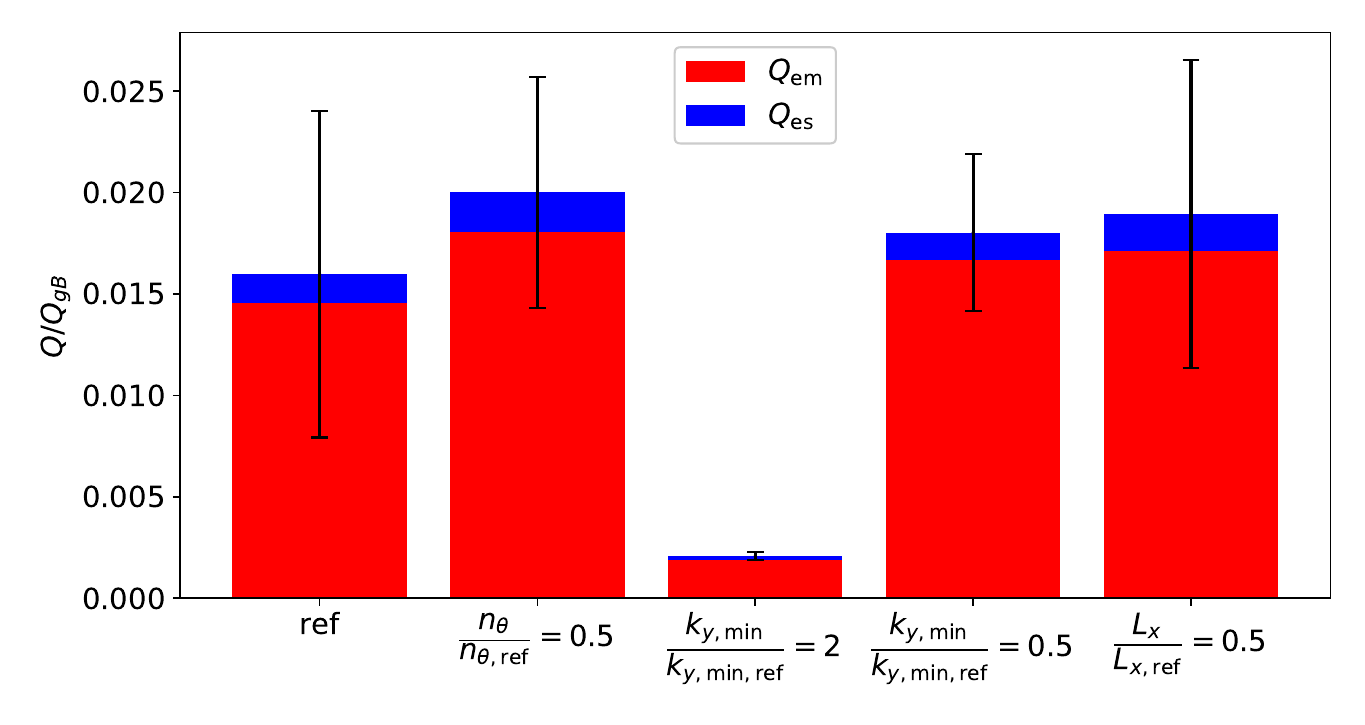}
    \caption{Saturated heat flux value from CGYRO nonlinear simulations at $\nu_e=0.42\,c_s/a$, $r/a=0.5$ and modified numerical resolution or domain size. The numerical resolution used in the reference nonlinear simulation is listed in table~\ref{tab:resolution_nl}. The error bar on the total heat flux is determined from standard deviation.}
    \label{fig:nl_res}
\end{figure}

\begin{figure}
    \centering
    \subfloat[]{\includegraphics[width=0.48\textwidth]{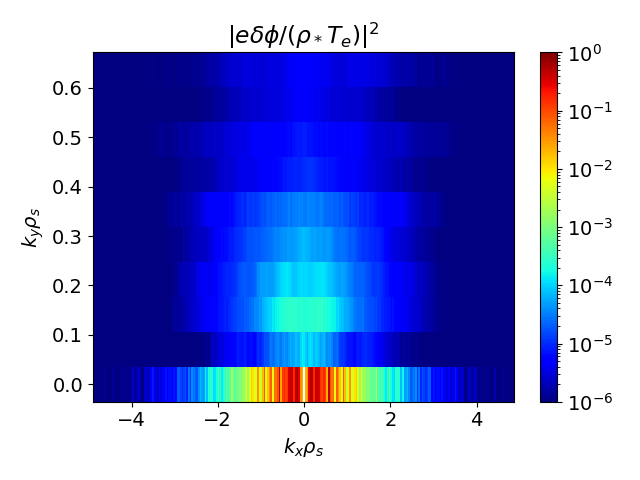}}\,
    \subfloat[]{\includegraphics[width=0.48\textwidth]{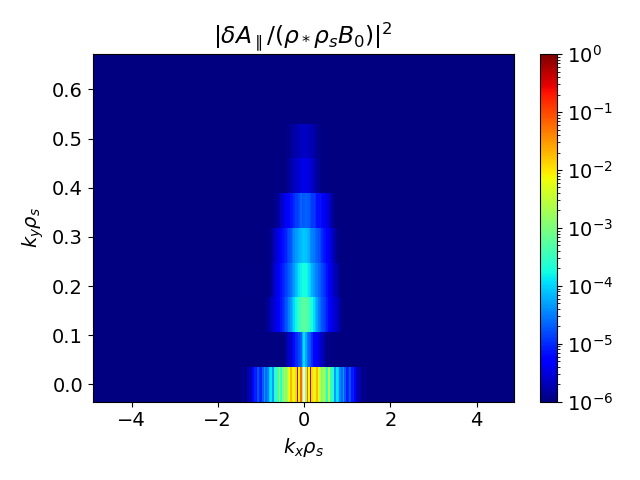}}
    \caption{Spectrum of $\delta \phi$  (a) and $\delta A_\parallel$ (b)  averaged over $t\in [1500, 2500]\,a/c_s$ and $\theta$ from the nonlinear simulation at $\nu_e = 0.42\,c_s/a$.}
    \label{fig:2Dspectra}
\end{figure}

The sensitivity to some numerical parameters is investigated by carrying out a set of nonlinear simulations with different parallel grid resolution, different values of $k_{y, \mathrm{min}}$ (the minimum finite $k_y$ mode evolved in the simulation) and different size of the radial flux-tube domain. 
In our nonlinear simulations $k_{y,\mathrm{min}}$ is varied to resolve the linearly unstable MTMs, but $k_\mathrm{y,\mathrm{max}}$ is kept constant.  The value of $k_{y,\mathrm{max}}$ is chosen to exclude the range of $k_y$ where ETG modes dominate (see Appendix D for further discussion).
The saturated heat flux level for each of these simulations is shown in Fig.~\ref{fig:nl_res}. The simulation with half the number of points in the parallel direction predicts the same heat flux value within the error bar. 
Also the simulation with $k_{y,\mathrm{min}} = k_{y,\mathrm{min, ref}}/2$ predicts the same heat flux level within the error bar. 
On the other hand, the simulation with $k_{y,\mathrm{min}} = 2k_{y,\mathrm{min, ref}}$ predicts a much lower heat flux. 
This is partially expected as modes at $k_y\rho_s\simeq 0.2$ are MTM unstable. The importance of low $k_y$ modes is highlighted in Fig.~\ref{fig:2Dspectra}, where $|\delta \phi(k_x, k_y)|^2$ and $|\delta A_\parallel(k_x, k_y)|^2$ spectra are shown as a function of $k_x$ and $k_y$. The maximum value of $|\delta \phi(k_x, k_y)|^2$ and $|\delta A_\parallel(k_x, k_y)|^2$ occurs at  $k_y\rho_s\simeq 0.15$ and small $k_x$ values. In the simulation with $k_{y,\mathrm{min}} = 2k_{y,\mathrm{min, ref}}$, the $|\delta \phi(k_x, k_y)|^2$ and $|\delta A_\parallel(k_x, k_y)|^2$ spectra are under resolved and the heat flux is therefore underestimated. 
We note that varying $k_{y,\mathrm{min}}$ may potentially affect the formation of stochastic layers due to magnetic island overlapping, as discussed in Sec.~\ref{sec:islands}.

\begin{figure}
    \centering
    \subfloat[]{\includegraphics[width=0.48\textwidth]{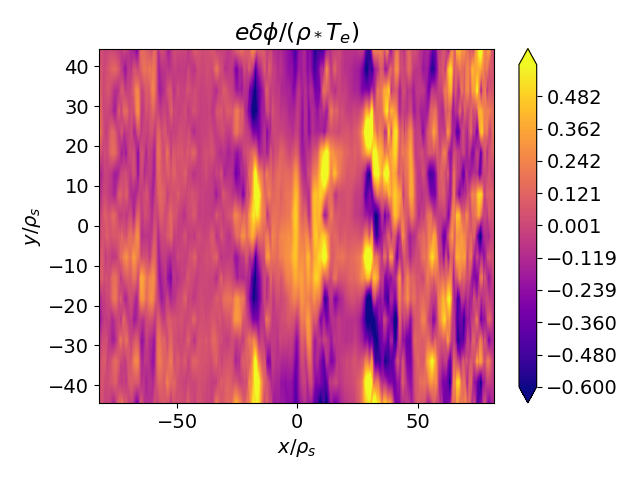}}\,
    \subfloat[]{\includegraphics[width=0.48\textwidth]{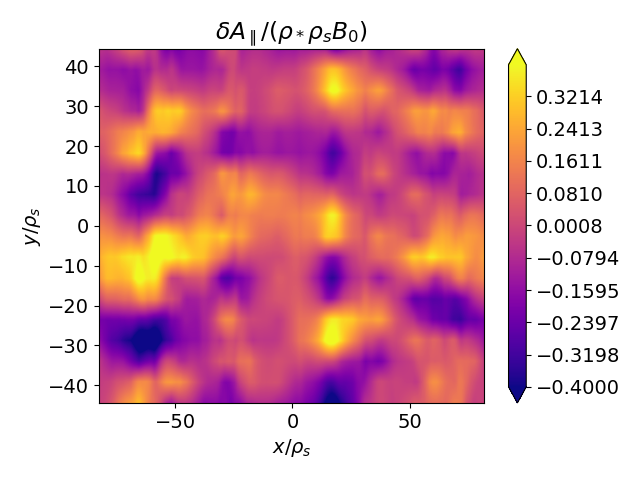}}
    \caption{Snapshot of $\delta \phi$ (a) and $\delta A_\parallel$ (b) from the nonlinear simulation with $\nu_e=0.42\,c_s/a$. Fluctuating quantities are shown without their zonal component.}
    \label{fig:2Dreal}
\end{figure}

The $\delta \phi$ spectrum peaks at low $k_y$ values and is extended in $k_x$ [see Fig.~\ref{fig:2Dspectra}~(a)], i.e. significant $\delta \phi$ amplitude is observed at high $k_x$. The electrostatic potential fluctuations are therefore very narrow radially and elongated in the binormal direction, as shown in Fig.~\ref{fig:2Dreal}~(a). 
On the opposite, the  $\delta A_\parallel$ spectrum is more narrow in $k_x$ [see Fig.~\ref{fig:2Dspectra}~(b)], thus resulting in much longer radial perturbations, as shown in Fig.~\ref{fig:2Dreal}~(b), which remain, however, smaller than the radial size of the flux tube domain.
Given the presence of elongated $\delta A_\parallel$ structures, the effect of the radial extent of the flux tube domain is tested. Fig.~\ref{fig:nl_res} shows that no significant difference on the heat flux level is observed in the simulation with $L_x = L_{x,\mathrm{ref}}/2$.
This is also expected from Fig.~\ref{fig:2Dreal}~(b), where the radial extent of $A_\parallel$ fluctuations is considerably smaller than the radial size of the flux tube domain, which therefore could be reduced by a factor of two without affecting  $\delta A_\parallel$. 
We note that, because of the low magnetic shear and $k_{y,\mathrm{min}}$ values, the $L_x$ and $L_y$ values used here are comparable to the size of MAST. This may question the applicability of the local approximation and global gyrokinetic simulations may be required to accurately predict the MTM driven heat flux, which is outside the scope of the present work in which we are interested in the saturation of MTMs within the local gyrokinetic framework.

\begin{figure}
    \centering
    \subfloat[]{\includegraphics[width=0.45\textwidth]{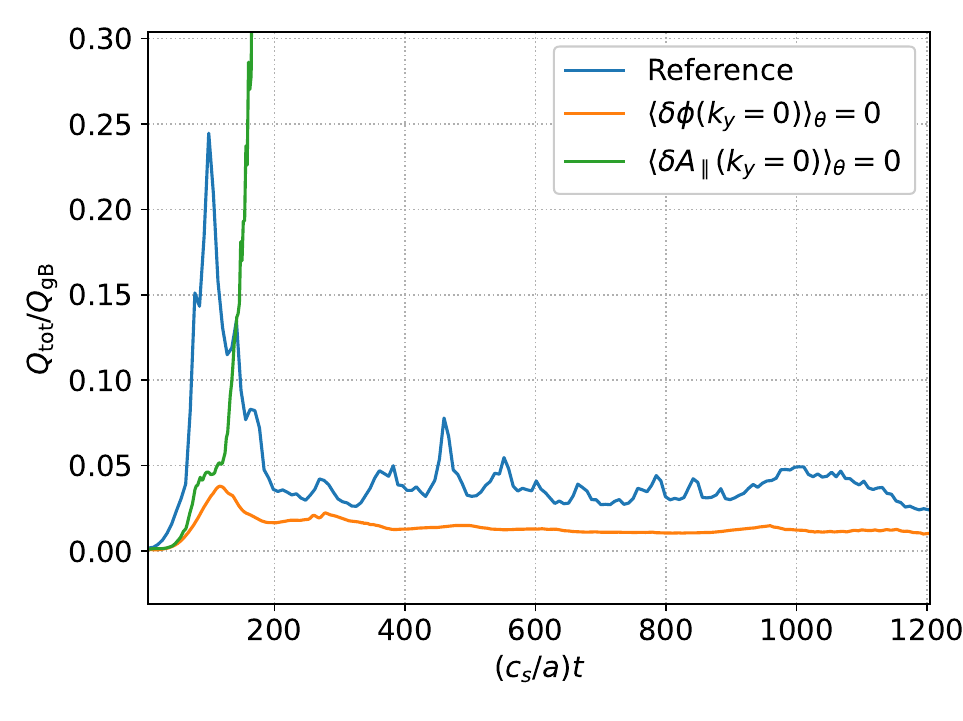}}\,
    \subfloat[]{\includegraphics[width=0.45\textwidth]{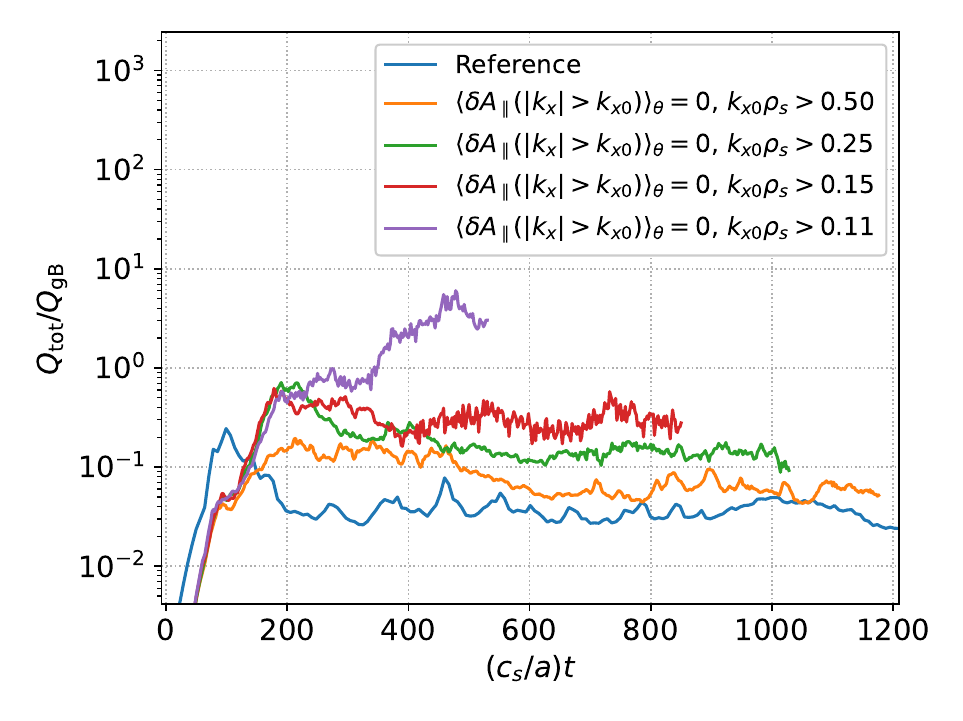}}\,
    \caption{(a) Time trace of the total heat flux from nonlinear tests where the zonal flows or zonal fields nonlinear interaction is turned off. The blue line represents the reference simulation ($\nu_e=0.42\, c_s/a$), the orange line represents a test with $\langle \delta \phi\rangle_\theta = 0$ and the red line a test $\langle \delta A_\parallel\rangle_\theta = 0$ in the nonlinear source term [see Eq.~(\ref{eqn:nlsource})]. (b) Nonlinear simulation tests with $\langle \delta A_\parallel(|k_x|>k_{x0})\rangle_\theta = 0$ in the nonlinear source term  for various values of $k_{x0}$.}
    \label{fig:zonal_test}
\end{figure}

The $|\delta \phi(k_x, k_y)|^2$ and $|\delta A_\parallel(k_x, k_y)|^2$ spectra in Fig.~\ref{fig:2Dspectra} reveal the presence of a very strong zonal component, which is more than an order of magnitude higher than the largest non zonal $\delta \phi$ and $\delta A_\parallel$ modes.
Zonal flows and/or zonal fields may therefore provide an important saturation mechanism here. In particular, the perturbed magnetic shear from zonal $A_\parallel$ perturbations, $\tilde{s} = qR/B(\mathrm{d}\delta B_y/\mathrm{d}x)\simeq 0.1$, is comparable to the local magnetic shear, $\hat{s}=0.34$, thus suggesting a potential important role of zonal fields. 
Saturation occurs via nonlinear interaction, where the nonlinear source term in the gyrokinetic equation can be written as~\cite{candy2016}
\begin{eqnarray}
S_{a, \mathrm{NL}} = \bigl[\chi_a, h_a\bigr] = \frac{\partial \chi_a}{\partial x}\frac{\partial h_a}{\partial y} - \frac{\partial h_a}{\partial x}\frac{\partial \chi_a}{\partial y}\,,
\label{eqn:nlsource}
\end{eqnarray}
where $x$ and $y$ are the radial and binormal coordinates, $h_a$ is the non adiabatic perturbed distribution function of species $a$, and $\chi_a$ is the generalised field potential,
\begin{eqnarray}
\chi_a = \biggl\langle \delta\phi(\mathbf{R}+\boldsymbol{\rho}) - v_\parallel \delta A_\parallel(\mathbf{R}+\boldsymbol{\rho}) - \mathbf{v}_\perp\cdot \delta\mathbf{A}_\perp(\mathbf{R}+\boldsymbol{\rho})\biggr\rangle_\mathbf{R}\,,
\end{eqnarray}
with $\mathbf{R}$ the guiding-center position, $\boldsymbol{\rho} = \mathbf{b}\times \mathbf{v}/\Omega_{ca}$ and $\langle \cdot\rangle_\mathbf{R}$ denoting the gyro-average (see Ref.~\cite{candy2016} for details). 
Different simulation tests are carried out to investigate the effect of $\langle\delta \phi(k_x, k_y=0)\rangle_\theta$ and $\langle\delta A_\parallel(k_x, k_y=0)\rangle_\theta$ in the nonlinear term of Eq.~(\ref{eqn:nlsource}). 
The time trace of the total heat flux from these tests is shown in Fig.~\ref{fig:zonal_test}. 
Removing $\langle\delta \phi(k_x, k_y=0)\rangle_\theta$ in Eq.~(\ref{eqn:nlsource}) has little effect on the saturated heat flux level, thus excluding any effect of zonal flows on the saturation mechanism. On the other hand, removing $\langle\delta A_\parallel(k_x, k_y=0)\rangle_\theta$ leads to a substantial increase of the heat flux, pointing out the main role played by zonal fields, in agreement with Ref.~\cite{pueschel2020}.

\begin{figure}
    \centering
    \subfloat[]{\includegraphics[width=0.45\textwidth]{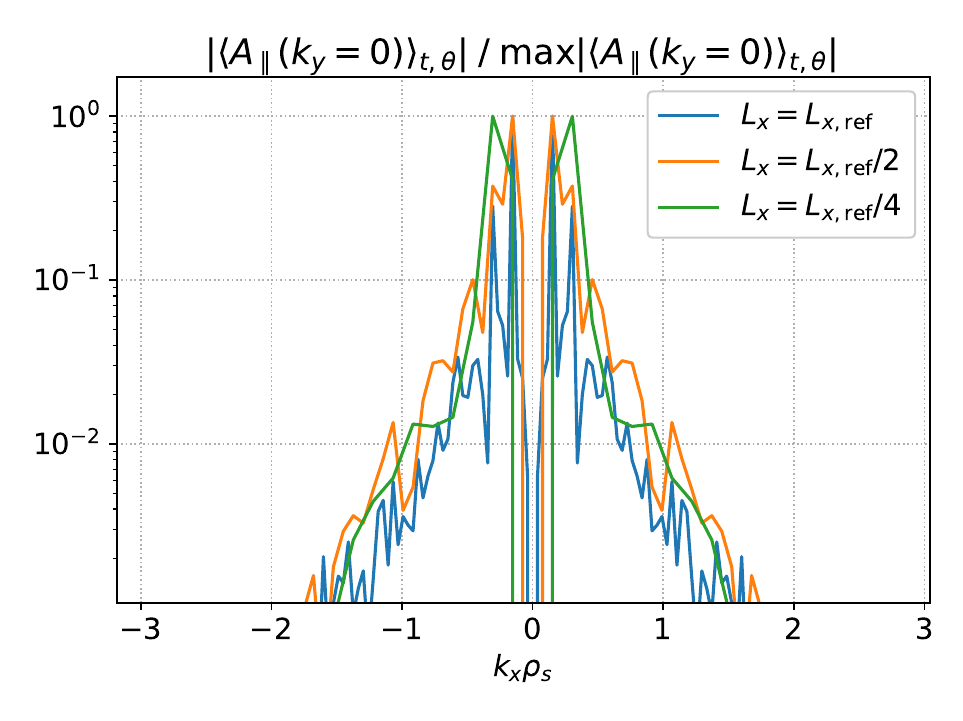}}\,
    \subfloat[]{\includegraphics[width=0.45\textwidth]{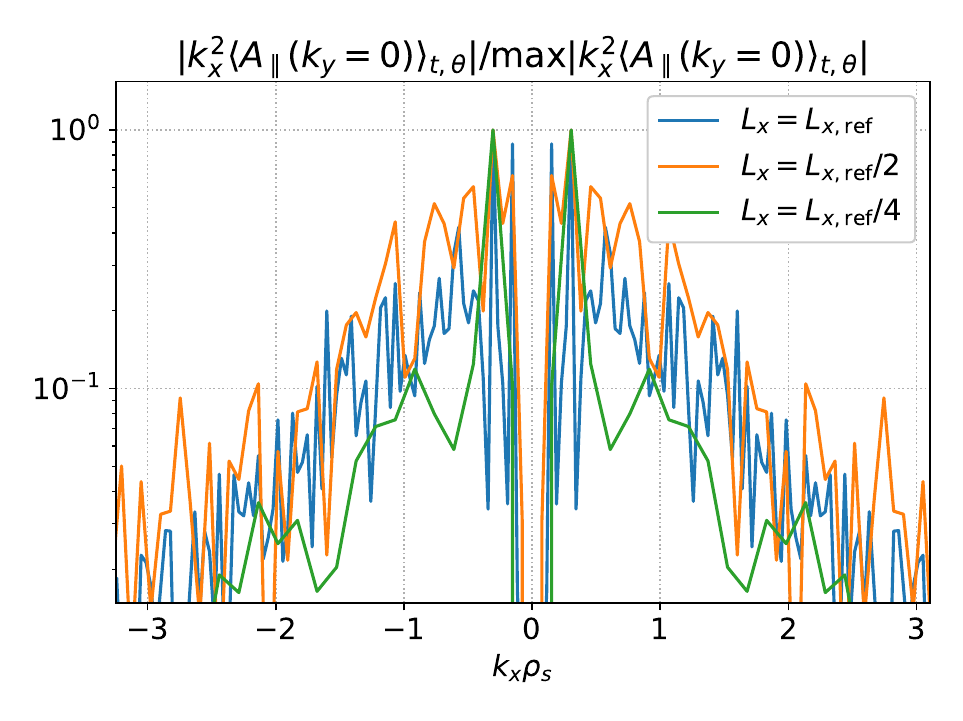}}\,
    \caption{Normalised zonal fields (a) and zonal fields shear (b) averaged over time and $\theta$ as a function of $k_x$ in the proximity of $k_x=0$ from nonlinear simulations at $\nu_e=0.42\,c_s/a$ with different radial domain extensions.}
    \label{fig:apar_zf}
\end{figure}

The $k_x$ spectrum of the zonal fields, $\langle A_\parallel(k_x, k_y=0)\rangle_\theta$, and their shear, $\langle k_x^2 A_\parallel(k_x, k_y=0)\rangle_\theta$, are shown in Fig.~\ref{fig:apar_zf} centered around $k_x=0$. We note that the zonal $A_\parallel$ spectrum peaks approximately at $|k_x\rho_s| \simeq 0.15$ and decays quickly as $|k_x|$ increases. The shear of the zonal fields shows a broader $k_x$ spectrum, with its maximum value occurring in the region between $|k_x\rho_s|\simeq 0.15$ and $|k_x\rho_s|\simeq 0.3$. Therefore, we expect an important role played by the low $k_x$ zonal field modes on the saturation mechanism. 
The effect of the low $k_x$ zonal field modes on the heat flux is tested in Fig.~\ref{fig:zonal_test}~(b) by removing $\langle \delta A_\parallel(k_x>k_{x0})\rangle_\theta$ in Eq.~(\ref{eqn:nlsource}) with different values of $k_{x0}$. 
Lowering $k_{x0}$ increases the damping of zonal fields and their associated shear, and the heat flux increases when $k_{x0}\rho_s$ is set below 0.25.
We note that a transition to very large heat flux values is observed only when the zonal field modes with $|k_{x0}\rho_s| > 0.11$ are removed from the nonlinear source term. 
We also mention that removing the zonal $A_\parallel$ modes with $|k_x\rho_s| \leq 0.11$ while retaining the modes with $|k_x\rho_s| > 0.11$ causes a transition to large heat flux values, thus confirming the important of the low $k_x$ zonal $A_\parallel$ modes.
Fig.~\ref{fig:apar_zf} shows that the maximum of $\langle \delta A_\parallel\rangle_\theta$ and $\langle k_x^2\delta A_\parallel\rangle_\theta$ spectra is well resolved in simulations with $L_x=L_{x, \mathrm{ref}}$ and $L_x=L_{x, \mathrm{ref}}/2$.
On the other hand, the resolution worsens in a simulation with $L_x=L_{x, \mathrm{ref}}/4 = 1/(s k_{y, \mathrm{min}})$, which is the minimum $L_x$ value that can be considered keeping all other parameters fixed. 
This simulation is affected by heat flux oscillations and convergence difficulties. 
The $k_x$ resolution is therefore important here to correctly capture the saturation mechanism via zonal fields.

\section{Magnetic islands interaction and local shear effect}
\label{sec:islands}

\begin{figure}
    \centering
    \includegraphics[scale=0.75]{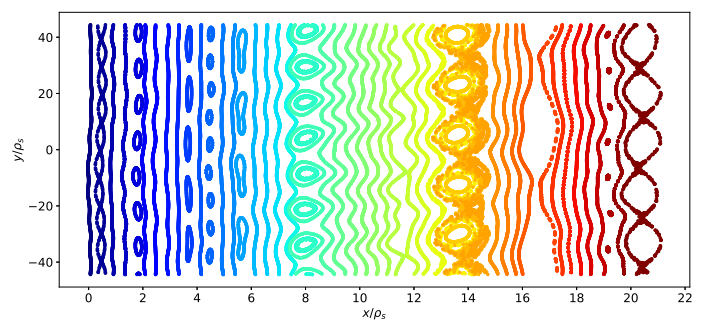}
    \caption{Poincaré map tracing where the magnetic field lines crosses the outboard midplane as they wind around the torus. This is obtained from the nonlinear simulation with $\nu_e = 0.42\ c_s/a$. The color scale is used to identify magnetic field lines starting at the same radial position. For the sake of clarity, only a part of the flux tube radial domain is shown.}
    \label{fig:poincare}
\end{figure}

\begin{figure}
    \centering
    \subfloat[]{\includegraphics[width=0.45\textwidth]{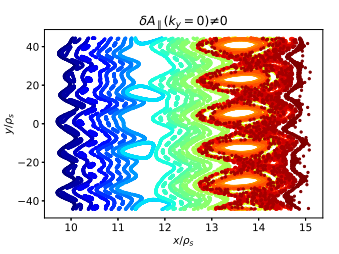}}
    \subfloat[]{\includegraphics[width=0.45\textwidth]{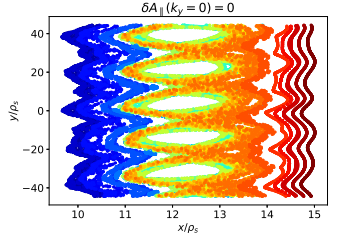}}
    \caption{Poincaré map of a thin radial layer of the flux tube domain generated from the nonlinear simulation with $\nu_e = 0.42\ c_s/a$. The zonal $\delta A_\parallel$ component is retained in (a) and turned off in (b) when generating the Poincaré map. The color scale is used to identify magnetic field lines starting at the same radial position.}
    \label{fig:poincare_zoom}
\end{figure}

We analyse here the effect of magnetic islands overlapping and subsequent formation of a stochastic layer. 
Fig.~\ref{fig:poincare} shows a Poincaré map of the magnetic field on a radial section of the flux tube domain generated from the reference nonlinear simulation with $\nu_e = 0.42\ c_s/a$.
Several magnetic islands with different mode numbers can be clearly distinguished. The largest island width corresponds to the mode at $k_y/k_{y,\mathrm{min}} = 2$, in agreement with the $\delta A_\parallel$ spectrum shown in Fig.~\ref{fig:2Dspectra}, where the largest non-zonal $\delta A_\parallel$  amplitude is achieved at $k_y/k_{y,\mathrm{min}} = 2$. The island width is smaller at higher $k_y$ and the subsequent perturbation of the equilibrium magnetic field is weaker. 
Since the minimum radial separation between adjacent islands is proportional to $1/n^2\propto 1/k_y^2$ (see Eq.~\ref{eqn:sep}), the effect of these high-$k_y$ islands can nonetheless be important.
For example, the radial separation of resonant surfaces for the $k_y/k_{y,\mathrm{min}}=5$ islands is $\Delta r = 1/(\hat{s} k_y) \simeq 8.4\ \rho_s$ and these islands appear in Fig.~\ref{fig:poincare} at $x \simeq 4\ \rho_s$, $x \simeq 13\ \rho_s$ and $x \simeq 21\ \rho_s$. The $k_y/k_{y, \mathrm{min}}=7$ islands appear at $x \simeq 2.2\,\rho_s$,  $x \simeq 8\,\rho_s$, $x \simeq 14\ \rho_s$ and $x \simeq 19.6\ \rho_s$ (not clearly visible in Fig.~\ref{fig:poincare}). The $k_y/k_{y,\mathrm{min}}=3$ and $k_y/k_{y,\mathrm{min}}=9$ islands also appear at a radial surface near $x=14\ \rho_s$. 
The overlap of these magnetic islands generates a layer of stochastic magnetic field lines in proximity of $x=14\ \rho_s$, as shown in Fig.~\ref{fig:poincare_zoom}~(a). We note that the stochastic layer around $x=14\ \rho_s$ is quite narrow in the radial direction and is surrounded by regions of weakly perturbed magnetic field.
In fact, the overall magnetic field stochasticity is relatively low, with regions of well separated islands and almost unperturbed magnetic field. 
This is consistent with the MTM driven heat flux at $\nu_e=0.42\ c_s/a$ (where the MTM instability is linearly most unstable) saturating at a relatively small value.

The main role of zonal fields is to saturate the non zonal $\delta A_\parallel$ at low amplitude, thus reducing the width of the magnetic islands and therefore the stochastic layer size (see Sec.~\ref{sec:nonlinear}).
A strong zonal $A_\parallel$ can directly affect the formation of a stochastic layer. This is shown in Fig.~\ref{fig:poincare_zoom}~(b), where the Poincaré map is generated without including the zonal $A_\parallel$ (the nonlinear simulation does include the zonal $A_\parallel$). By comparing Fig.~\ref{fig:poincare_zoom}~(a) and (b), we note that the magnetic islands at $x\simeq 12\ \rho_s$ and $x\simeq 14\ \rho_s$ merge together in the case without zonal $A_\parallel$, thus extending the radial size of the stochastic layer. We also note that the zonal $A_\parallel$ slightly shifts the radial position of rational surfaces. For example, the resonant surface with $k_y/k_{y,\mathrm{min}}=5$ moves from $x\simeq 12.5\ \rho_s$ to $x\simeq 13.5\ \rho_s$ by adding the zonal $A_\parallel$.

\begin{figure}
    \centering
    \subfloat[$\hat{s}=0.34$]{\includegraphics[width=0.45\textwidth]{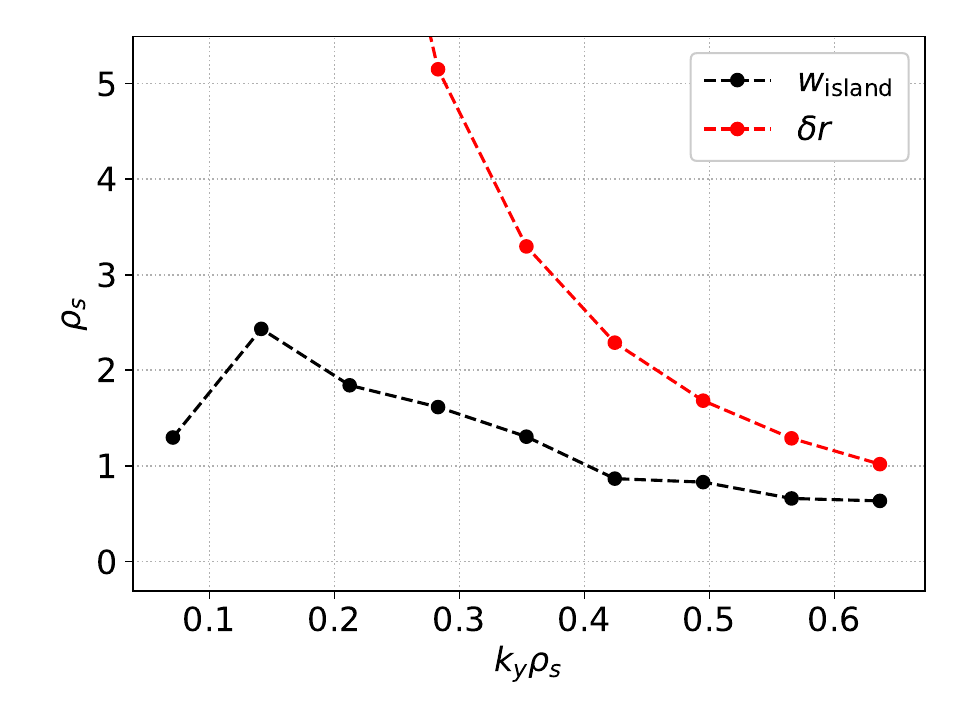}}\,
    \subfloat[$\hat{s}=0.7$]{\includegraphics[width=0.45\textwidth]{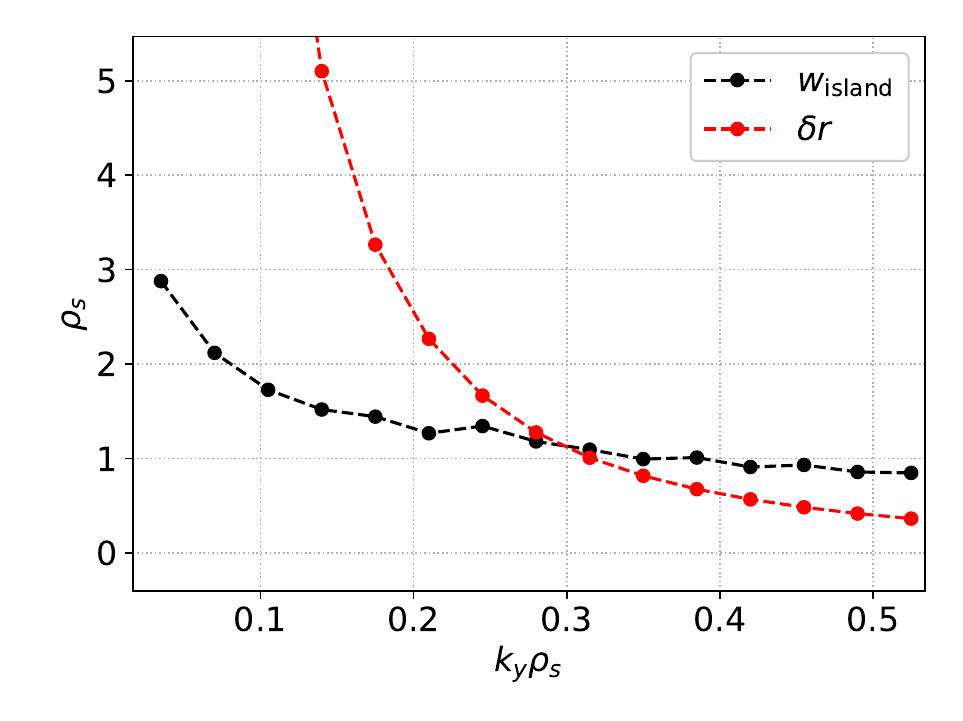}}
    \caption{Minimum resonant radial surface spacing (red line) and island width (black line) as a function of $k_y$ from the nonlinear simulations with $\hat{s}=0.34$ (a) and $\hat{s}=0.7$ (b).}
    \label{fig:wisland}
\end{figure}

Heat flux transport caused by stochastic magnetic field lines depends on both the size of magnetic islands and the separation between adjacent resonant surfaces. 
The island width is estimated from $\delta A_\parallel$ using Eq.~(\ref{eqn:wisland}), while the minimum separation between adjacent resonant surfaces is given by Eq.~(\ref{eqn:sep}). 
By following Refs.~\cite{nevins2011,guttenfelder2011}, we compare in Fig.~\ref{fig:wisland}~(a) $w_\mathrm{island}$ and $\delta r$ from the nonlinear simulation at $\nu_e=0.42\ c_s/a$.
We note that $\delta r$ is larger than $w_\mathrm{island}$ at all the $k_y$ modes evolved in the nonlinear simulations, in agreement with the low level of field lines stochasticity shown in Fig.~\ref{fig:poincare}. 

\begin{figure}
    \centering
    \subfloat[]{\includegraphics[width=0.45\textwidth]{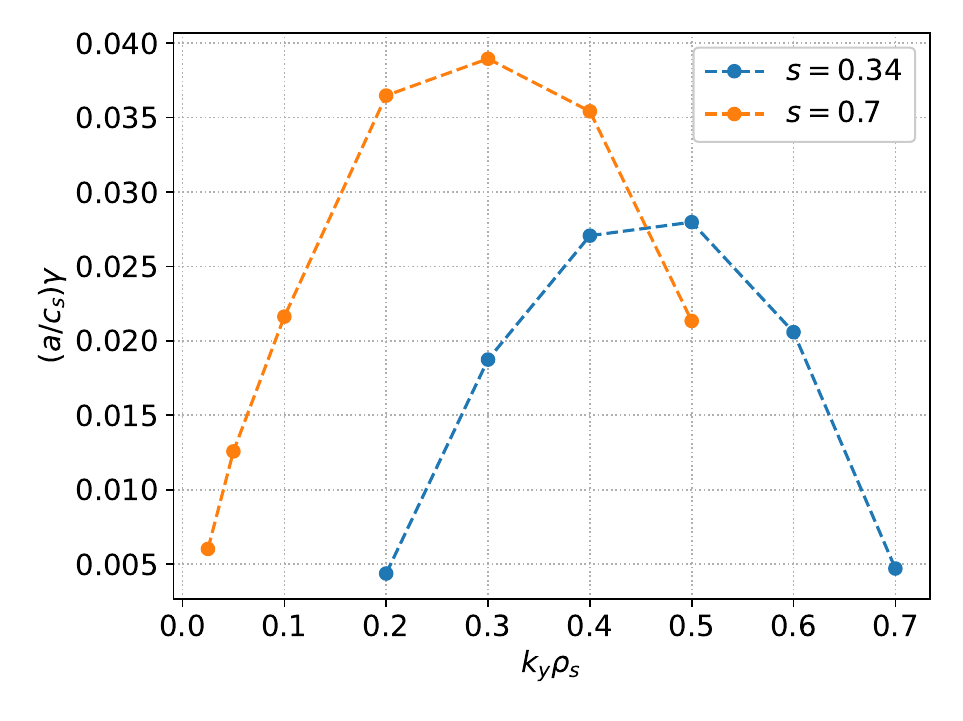}}\,
    \subfloat[]{\includegraphics[width=0.45\textwidth]{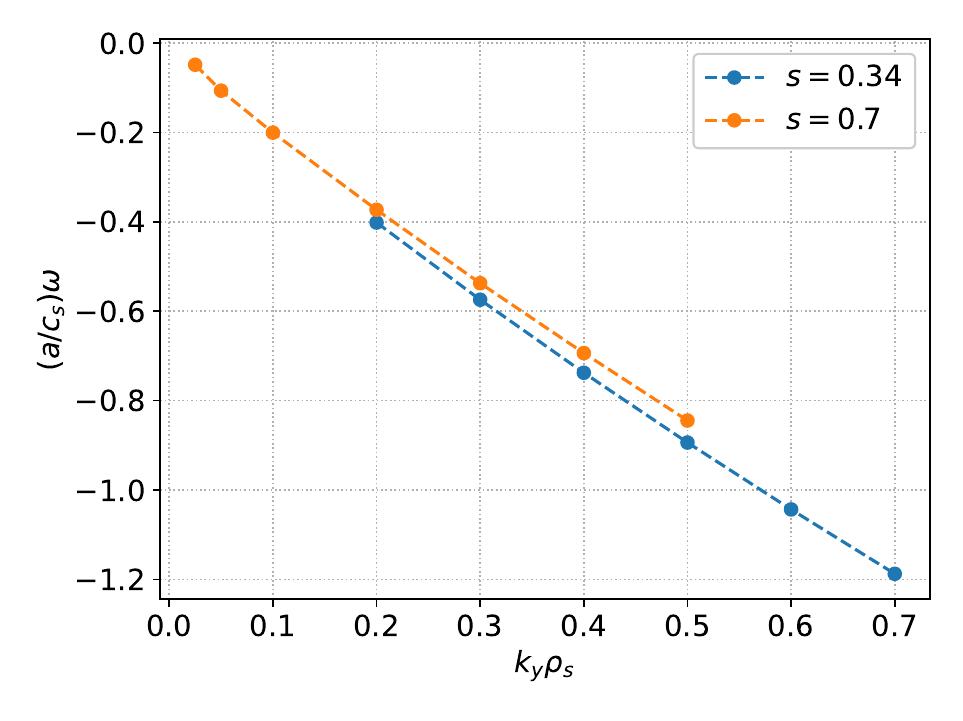}}
    \caption{Growth rate (a) and mode frequency (b) values from CGYRO linear simulations with $\hat{s}=0.34$ (blue line) and $\hat{s}=0.7$ (orange line). Only unstable modes are shown.}
    \label{fig:sh_lin}
\end{figure}

\begin{figure}
    \centering
    \includegraphics[scale=0.7]{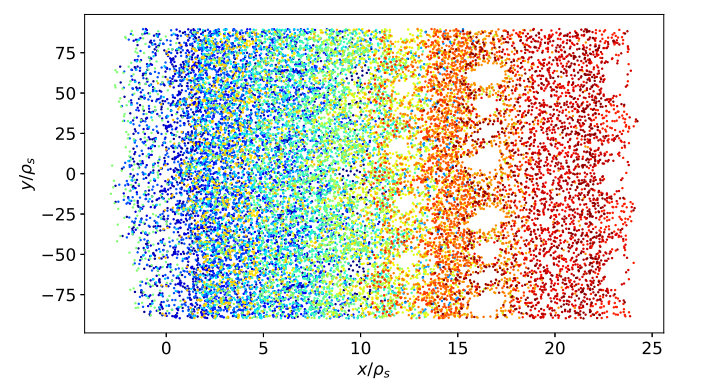}
    \caption{Poincaré map of magnetic field obtained from the nonlinear simulation with $\nu_e = 0.42\ c_s/a$ and higher magnetic shear. The color scale is used to identify magnetic field lines starting at the same radial position.}
    \label{fig:poincare_s07}
\end{figure}

\begin{figure}
    \centering
    \includegraphics[scale=0.5]{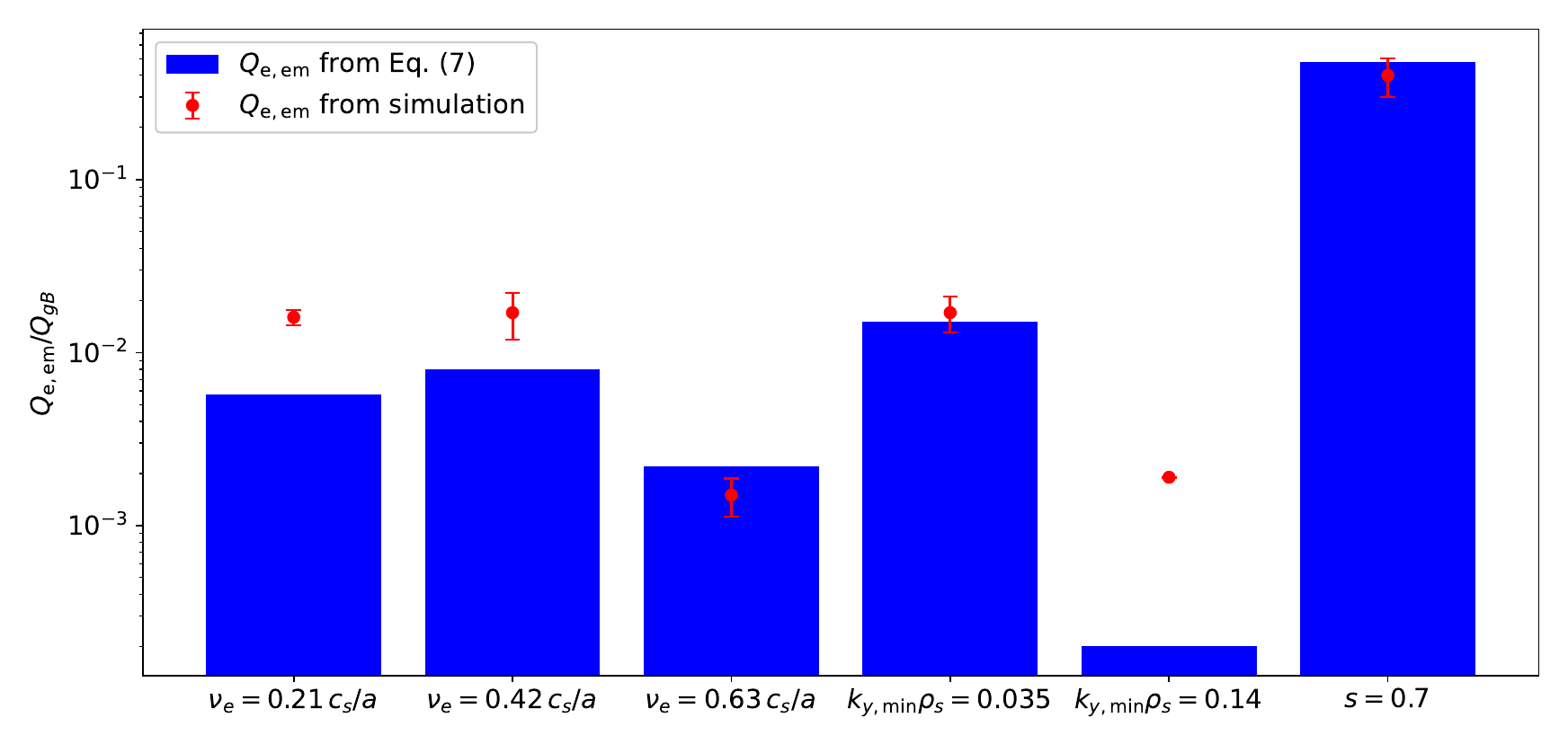}
    \caption{Electromagnetic electron heat flux due to magnetic diffusivity (see Eq.~(\ref{eqn:q_st})) in various nonlinear simulations with different values of electron collision frequency and $k_{y,\mathrm{min}}$. A nonlinear simulation with higher magnetic shear ($\hat{s}=0.7$) is also considered. The red markers show the saturated electromagnetic electron heat flux value computed from  nonlinear simulations. }
    \label{fig:stoc_flux}
\end{figure}

The resonant radial surface spacing is inversely proportional to the magnetic shear. A larger stochastic layer is therefore expected to form at higher magnetic shear. We consider therefore an additional case at higher magnetic shear, $\hat{s} = 2 \hat{s}_\mathrm{ref} \simeq 0.7$. 
The growth rate as a function of $k_y$ in this new case is shown in Fig.~\ref{fig:sh_lin}. The maximum growth rate of the MTM instability is higher than in the reference case with $\hat{s}=0.34$ and it occurs at a lower $k_y$ value. 
A nonlinear simulation at higher $\hat{s}$ is carried out with the numerical resolution listed in  table~\ref{tab:resolution_nl}. A lower value of $k_{y, \mathrm{min}}$ is used here to account for the low $k_y$ unstable modes. 
Fig.~\ref{fig:wisland}~(b) shows the comparison between $w_\mathrm{island}$ and $\delta r$ in the simulation at higher $\hat{s}$. The island width at different $k_y$ values is comparable to the one from the nonlinear simulation at $\hat{s}=0.34$. In fact, the amplitude of $A_\parallel$ fluctuations from the two simulations with different magnetic shear is comparable. 
This is in agreement with the nonlinear MTM theory developed in Ref.~\cite{drake1980}, which predicts $|\delta B/B| \simeq \rho_e/L_{T_e} \simeq 3.5\times 10^{-5}$ (the temperature gradient is the same in the two simulations). This value is close to the one obtained from  nonlinear simulations at $\nu_e = 0.42\ c_s/a$, i.e. $|\delta B/B| \simeq 6 \times 10^{-5}$.
On the other hand, the resonant surface spacing is a factor two smaller in the simulation at higher $\hat{s}$. This leads to an important qualitative change in Fig.~\ref{fig:wisland}~(b), where  $w_\mathrm{island}$ is larger than $\delta r$ for $k_y\rho_s>0.3$.
Consequently, island overlapping is more effective and a stochastic layer is expected to extend over a wider region.
This is clearly shown by the Poincaré map in Fig.~\ref{fig:poincare_s07}. The radial extent of the stochastic layer is much larger than in the reference case (see Fig.~\ref{fig:poincare}). Most of the magnetic islands visible in Fig.~\ref{fig:poincare} in the low magnetic shear simulation are destroyed in the high magnetic shear simulation by the presence of a stochastic layer. 
Consequently, the heat flux increases in the higher magnetic shear simulation (see Fig.~\ref{fig:stoc_flux}) and largely overcomes the heat flux driven by the ETG instability.
We note that $w_\mathrm{island}$ values in the low and high magnetic shear simulations are similar and the formation of a stochastic layer is mainly caused by a factor of four reduction of the minimum adjacent resonant surface spacing.

Following Ref.~\cite{rechester1978}, a magnetic diffusion coefficient is introduced,
\begin{eqnarray}
\label{eqn:dm}
D_m = \lim_{l\to\infty}\frac{\langle[r(l) - r(0)]^2\rangle}{2l} \simeq \lim_{l\to\infty}\frac{1}{2l}\frac{1}{N}\sum_{i=1}^{N}[r_i(l) - r_i(0)]^2\,,
\end{eqnarray}
where $r_i$ is the radial position of a field line, $l$ is the distance along the field line and $N$ is the number of field lines considered in the average. The magnetic diffusion coefficient converges to a well-defined value at large $N$. The magnetic diffusivity is computed in all the nonlinear simulations by considering the full radial extent, $N=400$ field lines and integrating along the perturbed field line for 2000 poloidal cycles. 
An estimate of the electron heat transport due to stochastic magnetic field lines can be derived from $D_m$~\cite{nevins2011},
\begin{eqnarray}
\label{eqn:q_st}
\frac{Q_{e,\mathrm{stochastic}}}{Q_{gB}} = 2 f_p\sqrt{\frac{4 m_i}{\pi m_e}}\frac{a}{L_{T_e}} \biggl(\frac{aD_m}{\rho_s^2}\biggr)\,
\end{eqnarray}
where $f_p \simeq 1-\sqrt{r/R}$ is the fraction of passing particles (magnetically trapped particles do not contribute to the stochastic transport~\cite{rechester1978}).
Fig.~\ref{fig:stoc_flux} compares the electron heat flux given by Eq.~(\ref{eqn:q_st}) to the electromagnetic electron heat flux calculated from nonlinear simulations. The trend is well reproduced and the predictions of Eq.~(\ref{eqn:q_st}) are in qualitative agreement with the heat flux predicted by nonlinear simulations. 
Importantly, the order of magnitude increase in the heat flux observed at higher magnetic shear is reproduced by Eq.~(\ref{eqn:q_st}).
We note that the electromagnetic electron heat flux is entirely due to the stochastic magnetic diffusivity in the case of $\hat{s}=2\hat{s}_\mathrm{ref}\simeq 0.7$, while a smaller contribution from the stochastic transport is observed in the simulations with the nominal magnetic shear value, though the stochastic contribution remains nonetheless important.

The minimum spacing between resonant surfaces depends on $n_0$ and, therefore, on $k_{y,\mathrm{min}}$ [see Eq.~(\ref{eqn:sep})].
As a consequence, different values of $k_{y,\mathrm{min}}$ may affect the stochastic layer and the subsequent saturated heat flux level. 
Fig.~\ref{fig:stoc_flux} shows the stochastic heat flux in the case with $k_{y,\mathrm{min}}/k_{y,\mathrm{min, ref}} = 2.0$ and $k_{y,\mathrm{min}}/k_{y,\mathrm{min, ref}} = 0.5$.
The value of $Q_{e,\mathrm{stochastic}}$ is much smaller when $k_{y,\mathrm{min}}/k_{y,\mathrm{min, ref}} = 2.0$ than in the reference case and it is significantly lower than the heat flux predicted by the nonlinear simulation.
As discussed in Sec.~\ref{sec:nonlinear}, a smaller heat flux is expected as the lowest $k_y$ unstable mode is not included in the simulation. 
The discrepancy between $Q_{e,\mathrm{stochastic}}$ and the electron heat flux from the simulation suggests that the stochastic magnetic diffusivity is underestimated by the simulation with $k_{y,\mathrm{min}}/k_{y,\mathrm{min, ref}} = 2$, and the (very low) heat flux is driven by a different mechanism. In fact, the minimum resonant surface spacing increases with $k_{y,\mathrm{min}}$, thus reducing the magnetic island overlap.
On the other hand, $Q_{e,\mathrm{stochastic}}$ at $k_{y,\mathrm{min}}/k_{y,\mathrm{min, ref}} = 0.5$ is similar to $Q_{e,\mathrm{stochastic}}$ at $k_{y,\mathrm{min}} = k_{y,\mathrm{min, ref}}$, despite the smaller resonant surface spacing. 
In fact, the amplitude of $\delta A_\parallel(k_y)$ decreases as the number of $k_y$ modes increases, thus preserving the condition $w_\mathrm{island} < \delta r$.
This comparison highlights that convergence on $k_{y,\mathrm{min}}$ should be carefully verified when performing nonlinear MTM simulations in order to avoid a possible underestimation of the stochastic heat flux, even when all the low $k_y$ unstable modes are included in the nonlinear simulation, especially when $w_\mathrm{island} \simeq \delta r$. 
The condition $w_\mathrm{island} \simeq \delta r$ can also be used to indicate when substantial stochastic electron heat flux might arise from MTMs.

\section{Gyrokinetic analysis at \texorpdfstring{$\boldsymbol{r/a=0.6}$}{TEXT}: experimentally significant \texorpdfstring{$\boldsymbol{Q_e^{\mathrm{MTM}}}$}{TEXT} at higher magnetic shear}
\label{sec:rad06}

In this section we briefly discuss the results of linear and nonlinear simulations carried out at the radial surface corresponding to $r/a=0.6$. The local parameters of this surface are shown in table~\ref{tab:equilibrium}. We note that this surface shares similar local parameter values with respect to the surface at $r/a=0.5$, extensively analysed in the previous sections, except for a higher value of magnetic shear, $\hat{s}=1.1$. Based on the results of Sec.~\ref{sec:islands}, we expect a more important role played by MTMs turbulence at $r/a=0.6$ than at $r/a=0.5$ (although we note that the value of $\beta$, important for MTMs, is slightly smaller at $r/a=0.6$ than at $r/a=0.5$). 

\begin{figure}
    \centering
    \includegraphics[scale=0.5]{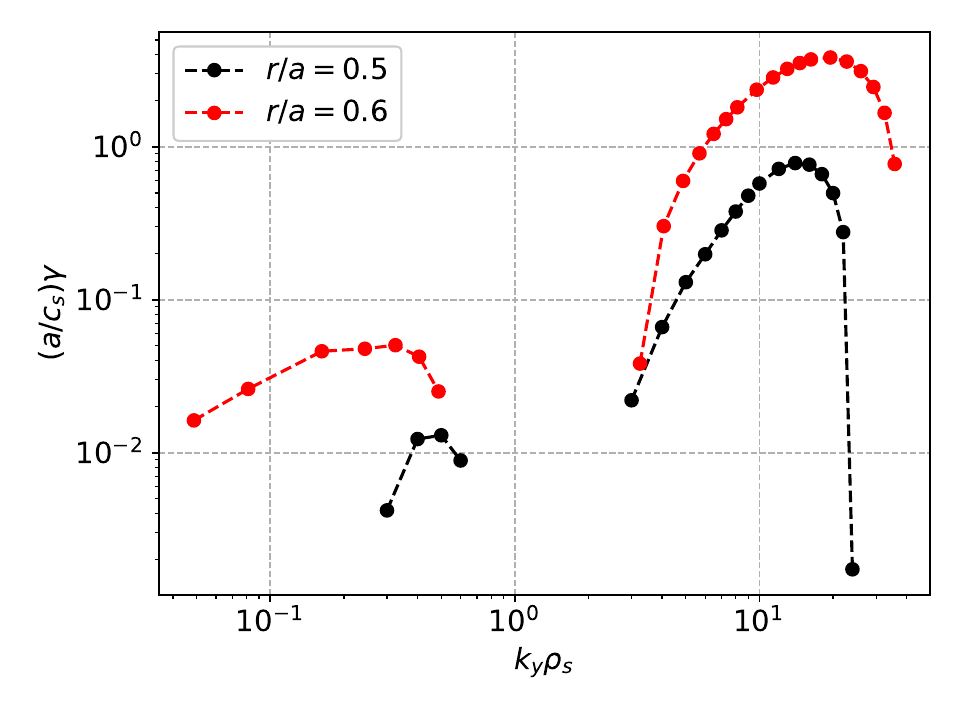}
    \caption{Growth rate as a function of $k_y$ at $r/a=0.5$ (black markers) and $r/a=0.6$ (red markers). The growth rate values are normalised to $c_s/a$, where $c_s$ is evaluated on the corresponding surface.}
    \label{fig:fullrad06}
\end{figure}

The growth rate values as a function of $k_y$ at $r/a=0.6$ are shown in Fig.~\ref{fig:fullrad06} along with the results at $r/a=0.5$ for comparison. MTMs and ETG modes dominate at ion and electron scale, respectively. We note that the maximum growth rate of both the MTM and ETG instabilities is larger at $r/a=0.6$ than at $r/a=0.5$. The MTM instability range extends at lower $k_y$ values at $r/a=0.6$. 
Despite these quantitative differences, the growth rate spectrum on the two surfaces is qualitatively similar.
 
\begin{figure}
    \centering
    \includegraphics[scale=0.5]{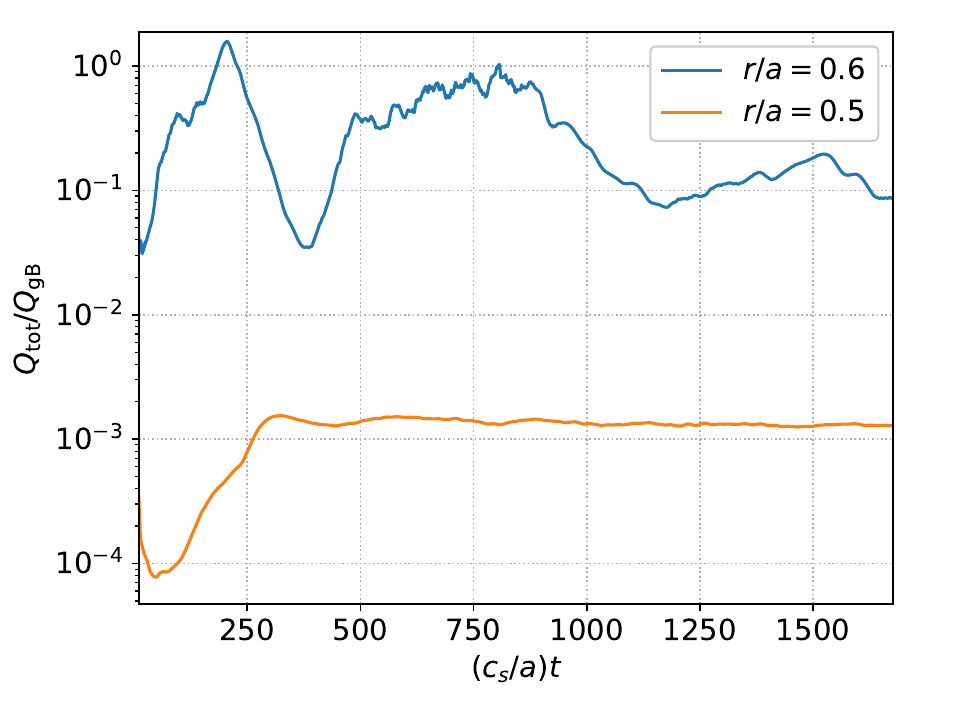}
    \caption{Time trace of the total heat flux driven by MTMs at $r/a=0.6$ (blue line) and $r/a=0.5$ (orange line). The flux is normalised to the gyro-Bohm value of the corresponding surface.}
    \label{fig:nlrad06}
\end{figure}

Similarly to Sec.~\ref{sec:linear}, we focus our analysis on the ion scale MTM instability. 
Fig.~\ref{fig:nlrad06} shows the time trace of the total heat flux, which is largely dominated by the magnetic flutter, from a nonlinear gyrokinetic simulation at $r/a=0.6$. For comparison purposes, Fig.~\ref{fig:nlrad06} shows also the total heat flux at $r/a=0.5$.
We note that the total heat flux driven by MTMs at $r/a=0.6$ saturates at approximately $Q_\mathrm{tot}\simeq 0.2\,Q_\mathrm{gB}$, corresponding to $Q_\mathrm{tot}\simeq 0.01$~MW/m$^2$, which is two orders of magnitude larger than the MTM-driven heat flux at $r/a=0.5$. 
We also remark that $Q_e^\mathrm{MTM}$ at $r/a=0.6$ is comparable to the ETG-driven heat flux (see~\ref{sec:etg}) and the total heat flux driven by MTMs and ETG modes is close to the experimental electron heat flux in this MAST discharge at $r/a=0.6$ (see Fig.~2(e) of Ref.~\cite{valovivc2011}).

The analysis at $r/a=0.6$ supports a role for MTMs complementing ETG modes in generating experimentally significant transport in higher $\hat{s}$ regions in MAST: MTMs together with ETG might therefore be expected to influence the observed scaling of energy confinement time with collisionality, $\tau_E \propto 1/\nu_*$~\cite{valovivc2011}, where candidate mechanisms to explain such a scaling have been proposed based on MTMs~\cite{guttenfelder2012} and ETG modes~\cite{colyer2017}.  We note, however, that our single scale computations of ETG and MTMs neglect multi-scale interactions between modes that may be important in setting the overall turbulent transport~\cite{maeyama2017}. However,  computationally demanding multi-scale nonlinear simulations are outside the scope of the present work.

\section{Conclusions}
\label{sec:conclusion}

The MTM instability can provide significant electron heat flux transport in the high-$\beta$ core of spherical tokamaks as well as in the edge of conventional aspect ratio tokamaks.
An accurate prediction of MTM-driven heat flux requires one to perform expensive nonlinear gyrokinetic simulations, which are often very challenging because of their numerical requirements and convergence difficulties.
This motivates improved understanding of the saturation and transport mechanisms to aid the development of cheaper reduced models.
In this work, the results of the first gyrokinetic nonlinear simulations of MTM turbulence in a MAST scenario are presented, therefore extending the linear analysis of Ref.~\cite{valovivc2011}. 
Linear simulations show that MTMs are the dominant linear instability at $k_y\rho_s < 1$ in all the cases considered here and that this MTM instability is sensitive to electron temperature gradient, density gradient, magnetic shear and electron collision frequency. 
At $r/a=0.5$ the magnetic shear is low ($\hat{s}=0.34$) and MTMs drive a negligible fraction of the total electron heat flux. On the other hand, slightly further out ($r/a=0.6$) the magnetic shear is larger ($\hat{s}=1.1$) and the MTM-driven heat flux is much larger and experimentally relevant. 
This MTM instability requires a velocity dependent electron collision frequency, it is weakly affected by ion dynamics or parallel magnetic fluctuations, and it is strongly destabilised by the inclusion of the electrostatic potential. 
A comparison between CGYRO and GS2 linear simulations is also carried out, showing an overall good agreement both in the reference case and in a case of stronger MTM drive.

The heat flux driven by MTMs at $r/a=0.5$ is negligible in the nonlinear simulation with the reference value of electron collision frequency as compared to the ETG driven heat flux, but reducing $\nu_e$ by a factor of two results in $Q_e^\mathrm{MTM}$ rising by an order of magnitude, but still only to a level where $Q_e^\mathrm{MTM} < 0.5\,Q_e^\mathrm{ETG}$.
A strong zonal $\phi$ and $A_\parallel$ is observed in all the nonlinear simulations. While the effect of zonal $\phi$ on the saturated level of heat flux is weak, a much larger heat flux is obtained when the zonal $A_\parallel$ is removed from the nonlinear source term, thus pointing out the importance of zonal fields in the saturation mechanism of this MTM instability in the considered MAST case. 

The MTM instability leads to the formation of magnetic islands at resonant surfaces. These magnetic islands can overlap if their width exceeds the radial separation between adjacent resonant surfaces, thus generating a layer of stochastic magnetic field lines. The effect of the stochastic layer formation and its radial extent on the heat flux is analysed in various nonlinear simulations. 
In the reference MAST case at $r/a=0.5$, the island widths are smaller than the minimum spacing between resonant surfaces and this is consistent with the saturation at low heat flux level. On the other hand, the heat flux increases by more than an order of magnitude when the value of the magnetic shear at $r/a=0.5$ is doubled. In this case, the island width exceeds the radial spacing between rational surfaces and a radially extended stochastic layer forms. 

Heat transport caused by stochastic magnetic field lines is quantified through the magnetic diffusivity $D_m$ given in Eq.~(\ref{eqn:dm}). A reasonable agreement is found between the electromagnetic heat flux predicted by nonlinear simulations and the stochastic heat flux at different values of electron collision frequency and magnetic shear, thus suggesting that the MTM driven heat flux in the cases considered here is mostly due to stochastic magnetic field diffusivity and confirming the important role played by the formation of stochastic layers.
The criterion $w_\mathrm{island} \gtrsim \delta r$ can therefore be used to indicate when substantial electron heat transport might be expected to arise from stochastic fields.
The saturated level of heat flux is shown to depend on the radial extent of the stochastic layer, which in turn depends on both the magnetic island width, proportional to the magnetic fluctuation amplitude, and the radial separation between adjacent resonant surfaces.
Nonlinear simulations with similar magnetic perturbation amplitude but different radial separation between adjacent resonant surfaces are shown to saturate at very different values.
It is also interesting to note that reduced Rechester-Rosenbluth based models of electron heat transport from stochastic magnetic fields~\cite{rechester1978} have been shown to model transport in MAST discharges reasonably well~\cite{palermo2022}.

Nonlinear simulations at $r/a=0.6$ at higher magnetic shear find much larger $Q_e^\mathrm{MTM}$ that is similar in magnitude to $Q_e^\mathrm{ETG}$  and experimentally relevant, thus supporting a role for MTM turbulence complementing ETG modes in generating experimentally significant transport in higher $\hat{s}$ regions of MAST.
Both MTMs and ETG modes might therefore be expected to influence the observed scaling of energy confinement time with collisionality, $\tau_E \propto 1/\nu_*$~\cite{valovivc2011}. 
We remark that our simulations do not account for the multi-scale interaction between ETG modes and MTMs. Multi-scale simulations are computationally very expensive and outside the scope of the present work. Future work will be performed in MAST scenarios to specifically address the multi-scale interaction between MTMs and ETG.

We also note that further work is required to derive a relation between magnetic fluctuation amplitude, adjacent resonant surfaces separation and saturated heat flux value, which will extend current quasi-linear theories (see, e.g., Ref.~\cite{rafiq2016}) and lead towards reliable and accurate MTM driven heat flux predictions from reduced models.

\section*{Acknowledgements}

The authors thank E. Belli, J. Candy, Ajay C. J., W. Guttenfelder and H. Wilson for useful discussions. 
This work has been supported by the Engineering and Physical Sciences Research Council (grant numbers EP/R034737/1 and EP/W006839/1).
Simulations have been performed on the ARCHER2 UK National Supercomputing Service under the project e607 and on the Viking research computing cluster at the University of York.

\appendix

\section{Linear convergence tests in the reference case}
\label{sec:convergence}

\begin{figure}
    \centering
    \subfloat[]{\includegraphics[width=0.47\textwidth]{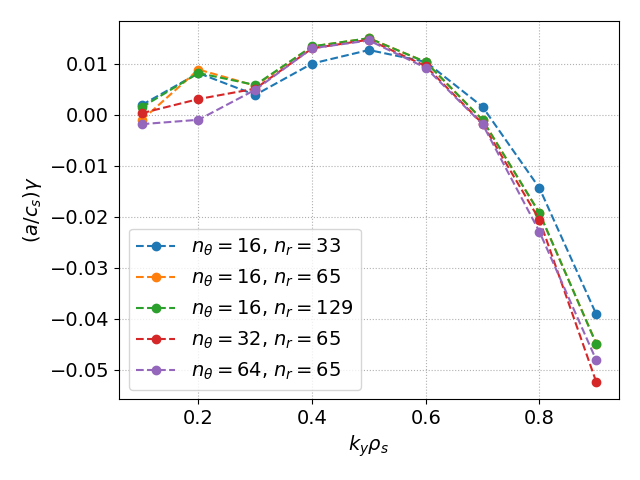}}\,
    \subfloat[]{\includegraphics[width=0.47\textwidth]{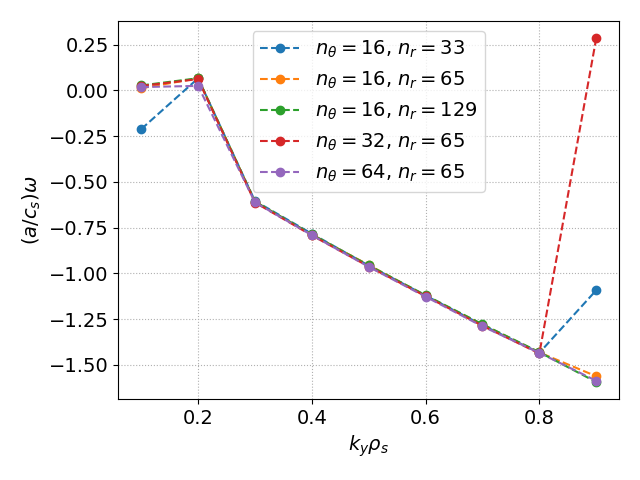}}
    \caption{Growth rate (a) and mode frequency (b) as a function of $k_y$ from GS2 linear simulations with $\delta B_\parallel=0$ at different values of $n_\theta$ and $n_r$. }
    \label{fig:lin_con}
\end{figure}

Linear simulations with different numerical resolution have been carried out in order to verify numerical convergence. Fig.~\ref{fig:lin_con} shows the growth rate and mode frequency from GS2 linear simulations performed at different values of $n_\theta$ and $n_r$. These tests have been performed without evolving $\delta B_\parallel$, which has a weak effect on the MTM instability (see Fig.~\ref{fig:ref}).
At $k_y\rho_s \geq 0.3$, convergence is achieved at $n_\theta \geq 32$ and $n_r\geq 65$, while the mode at $k_y\rho_s = 0.2$ requires a higher resolution along $\theta$. On the other hand, this mode is not driven unstable by the MTM instability and is stable when $\delta B_\parallel \ne 0$.

\section{ETG driven heat flux in the reference case}
\label{sec:etg}

\begin{figure}
    \centering
    \subfloat[$r/a=0.5$]{\includegraphics[width=0.47\textwidth]{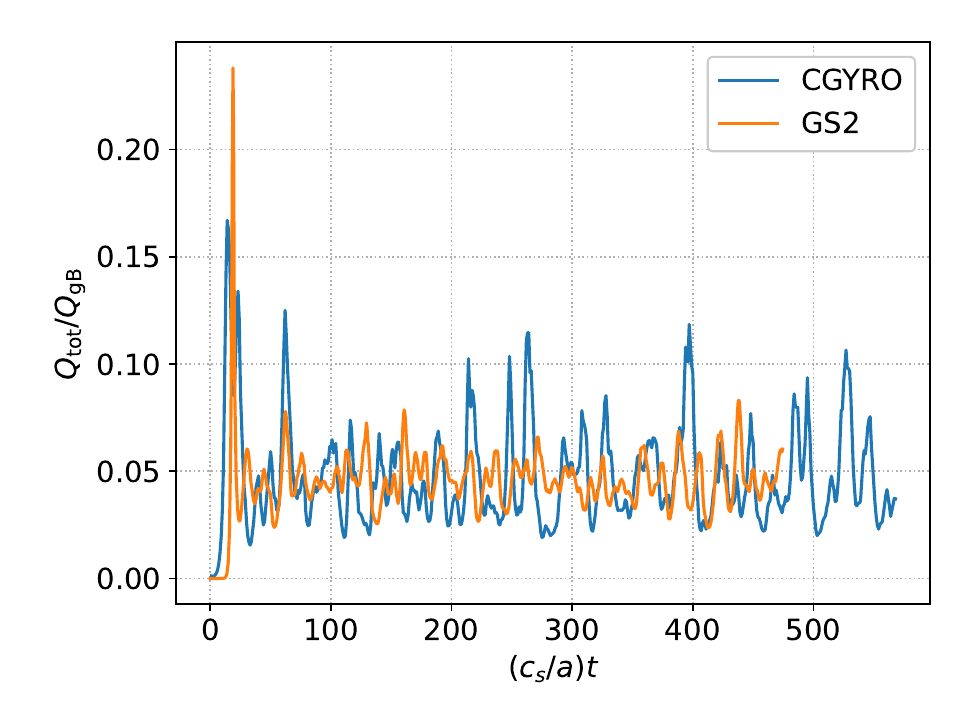}}\,
    \subfloat[$r/a=0.6$]{\includegraphics[width=0.46\textwidth]{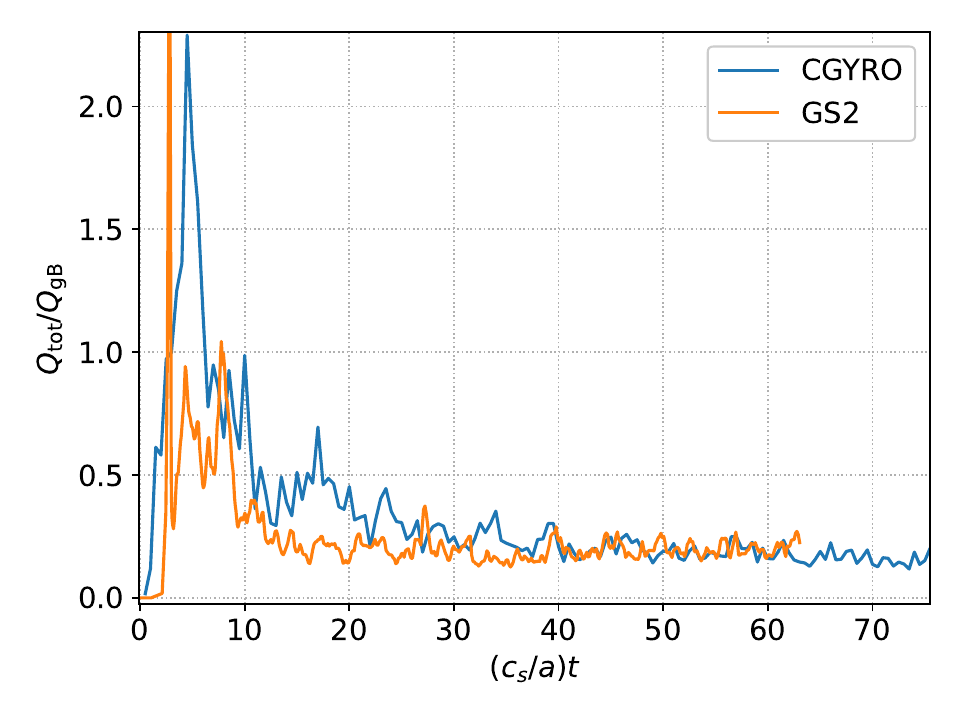}}\,
    \caption{Time trace of the total heat flux driven by the ETG instability in the reference MAST case at $r/a=0.5$ (a) and $r/a=0.6$ (b). Results from CGYRO (blue line) and GS2 (orange line) nonlinear simulations. The heat flux is normalised to the gyro-Bohm value of the corresponding surface.}
    \label{fig:qtot_etg}
\end{figure}

The linear gyrokinetic simulations described in Sec.~\ref{sec:reference} and Sec.~\ref{sec:rad06} show the presence of an ETG instability in the MAST reference case both at $r/a=0.5$ and at $r/a=0.6$. 
Since the aim of this work is to investigate ion scale collisional MTM turbulence in MAST, only ion scale instabilities is considered.
Here we show that the ETG-driven heat flux significantly contribute to the total heat flux both at $r/a=0.5$ (where it is the dominant transport mechanism) and at $r/a=0.6$. 
Fig.~\ref{fig:qtot_etg} shows the time trace of the total heat flux from GS2 and CGYRO nonlinear simulations of the reference MAST case at $r/a=0.5$ and $r/a=0.6$.
The heat flux saturates at $Q_\mathrm{tot}/Q_{gB} \simeq 0.05$ on the $r/a=0.5$ surface and at $Q_\mathrm{tot}/Q_{gB} \simeq 0.2$ on the $r/a=0.6$ surface, where $Q_{gB}$ is approximately 0.09~MW/m$^2$ at $r/a=0.5$ and 0.05~MW/m$^2$ at $r/a=0.6$. We note that the ETG-driven heat flux values are of the same order of magnitude of the heat flux computed by using TRANSP for this MAST discharge (see Fig.~2 of Ref.~\cite{valovivc2011}).
We highlight that the ETG heat flux is approximately two orders of magnitude larger than the MTM-driven heat flux at $r/a=0.5$ while it is comparable to the MTM-driven heat flux at $r/a=0.6$. 

\section{Binormal ion scale ETG instability}
\label{sec:etg_linear}

Fig.~\ref{fig:coll2D} shows that, when the electron collision frequency value is decreased, two different instabilities appear: an instability with positive mode frequency at $k_y\rho_s < 0.4$, which is identified as an ITG instability, and an instability with negative mode frequency at $k_y\rho_s > 0.4$, which we show here to be the electron scale ETG instability that extends into the ion scale $k_y$ region at low electron collision frequency.

\begin{figure}
    \centering
    \subfloat[]{\includegraphics[width=0.45\textwidth]{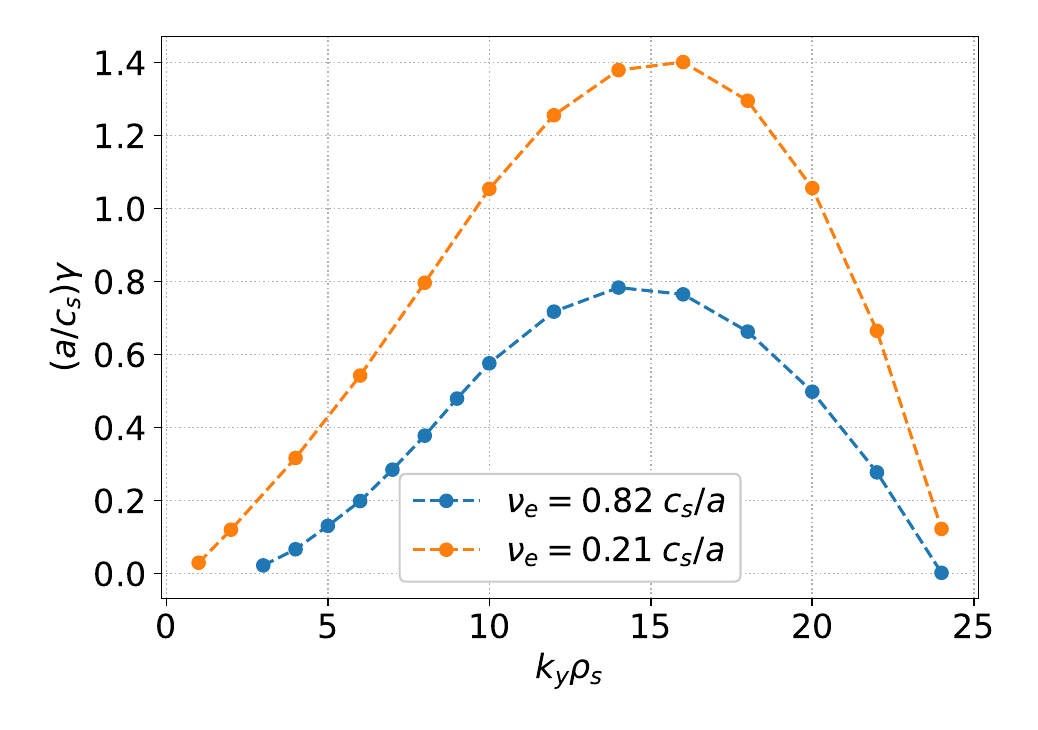}}\,
    \subfloat[]{\includegraphics[width=0.45\textwidth]{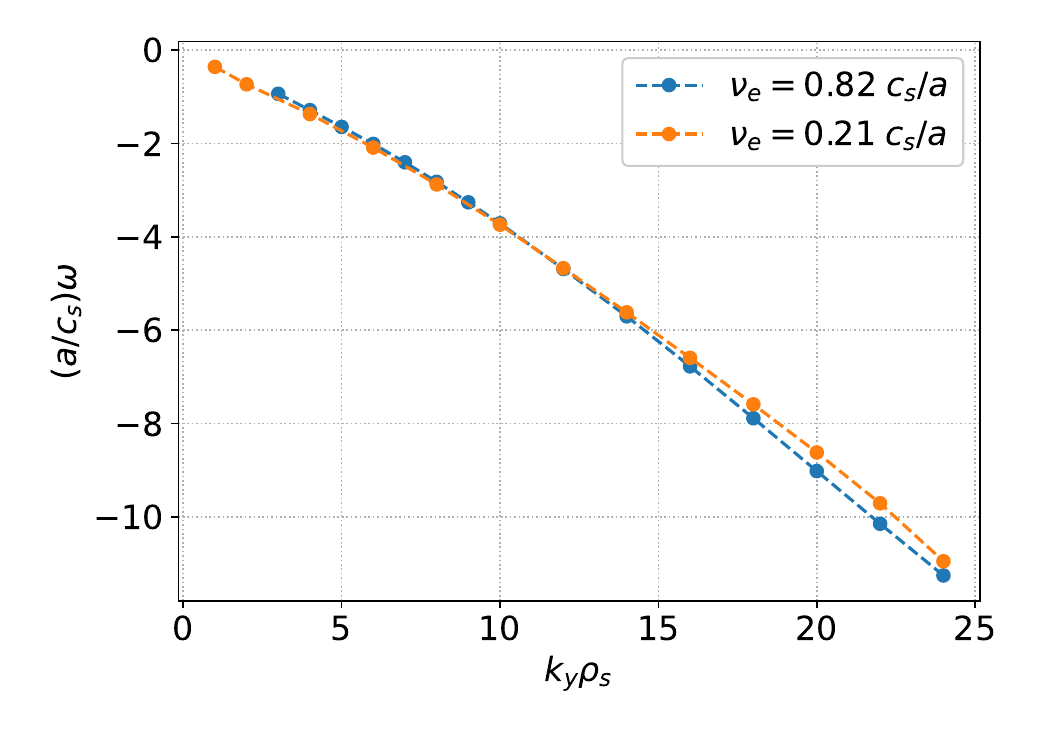}}
    \caption{Growth rate (a) and mode frequency (b) values from GS2 linear simulations at $k_y\rho_s>0.8$ with $\nu_e =0.82\,c_s/a$ (reference values) and $\nu_e =0.21\,c_s/a$. Only unstable modes are shown.}
    \label{fig:etg_lin}
\end{figure}

Fig.~\ref{fig:etg_lin} shows the growth rate and the mode frequency values of unstable modes with $k_y\rho_s > 0.8$ from GS2 linear simulations at the reference value of electron collision frequency and at a lower value. 
The growth rate of the ETG instability is higher at lower $\nu_e$ than in the reference case and the ETG instability extends to modes at $k_y\rho_s \simeq 1$. 

\begin{figure}
    \centering
    \subfloat[]{\includegraphics[width=0.45\textwidth]{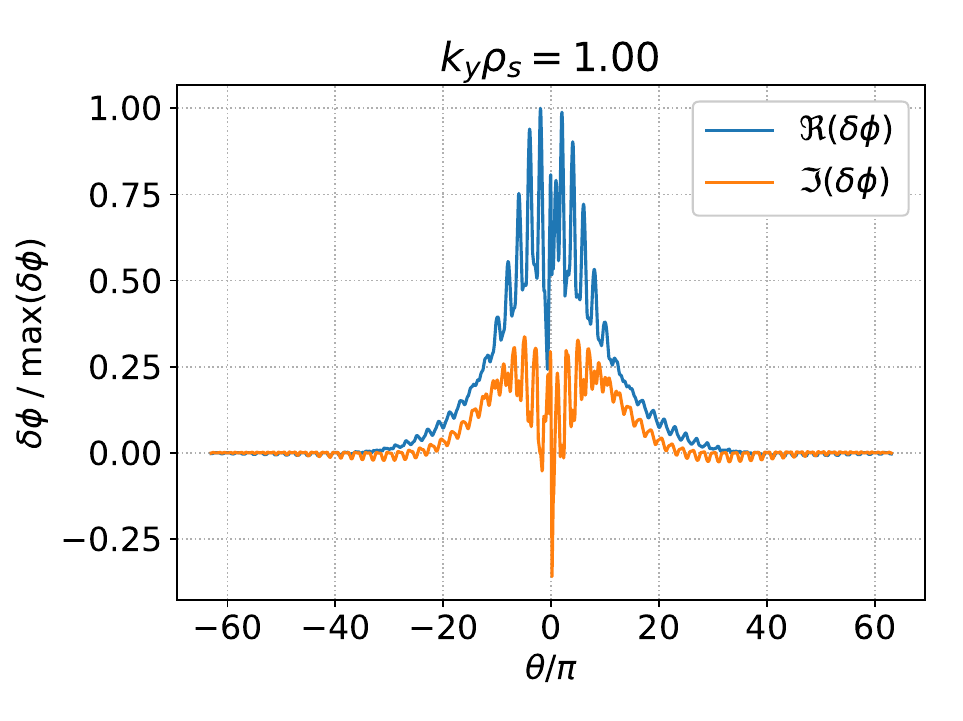}}\,
    \subfloat[]{\includegraphics[width=0.45\textwidth]{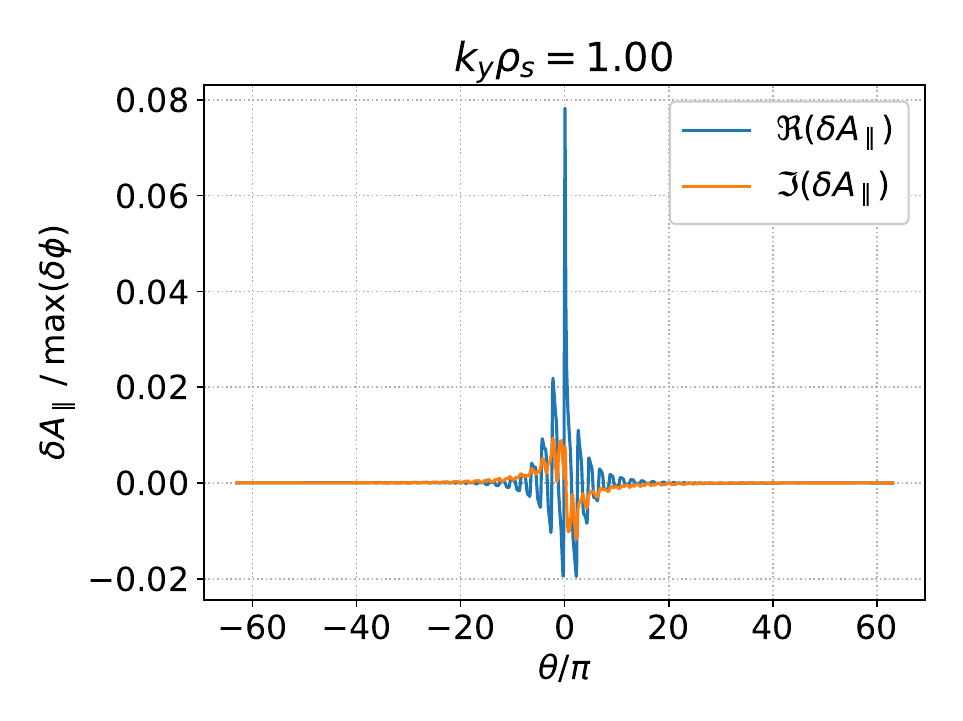}}
    \caption{Real and imaginary part of $\delta \phi/\max(\delta\phi)$ (a) and $\delta A_\parallel/\max(\delta\phi)$ (b) as a function of $\theta$ at $k_y\rho_s = 1.0$. Results from the GS2 linear simulation at $\nu_e = 0.21\,c_s/a$. The fields $\delta\phi$ and $\delta A_\parallel$ are normalised to $\rho_*T_e/e$ and $\rho_*\rho_sB_0$, respectively.}
    \label{fig:etg_eig_app}
\end{figure}

The real and imaginary parts of $\delta \phi$ and $\delta A_\parallel$ at $k_y\rho_s = 1.0$ are shown in Fig.~\ref{fig:etg_eig_app} for the simulation with $\nu_E=0.21\,c_s/a$.
Electrostatic potential mode structure is quite extended along $\theta$ and it is similar to the one in Fig.~\ref{fig:eig_etg}, which shows $\delta\phi(\theta)$ at $k_y\rho_s=0.5$ in the case of large electron temperature gradient values.
Therefore, while the MTM instability is suppressed at large $a/L_{T_e}$ values, the ETG instability extends into ion scale $k_y$ region, similarly to what is observed when the value of $\nu_e$ is decreased.

These ETG modes, which are unstable at $k_y\rho_s\simeq 1$ and very extended along the magnetic field line, are similar to the ones characterised in Ref.~\cite{hardman2022}.
We note that decreasing the collisionality decreases the electron detrapping frequency and therefore increases the drive at lower $k_y$ and $\omega$ values, as shown in appendix A of Ref.~\cite{roach2009}.

\section{Heat flux spectrum and dependence on \texorpdfstring{$\boldsymbol{k_{y,\mathrm{max}}}$}{TEXT}}
\label{sec:kymax}

\begin{figure}
    \centering
    \subfloat[]{\includegraphics[width=0.47\textwidth]{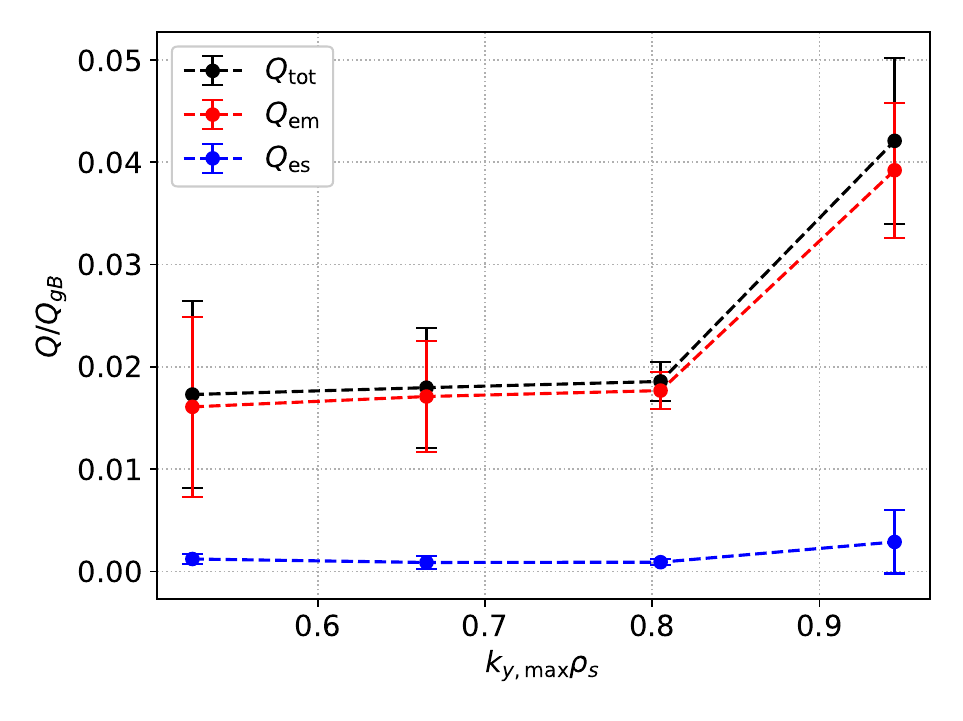}}\,
    \subfloat[]{\includegraphics[width=0.49\textwidth]{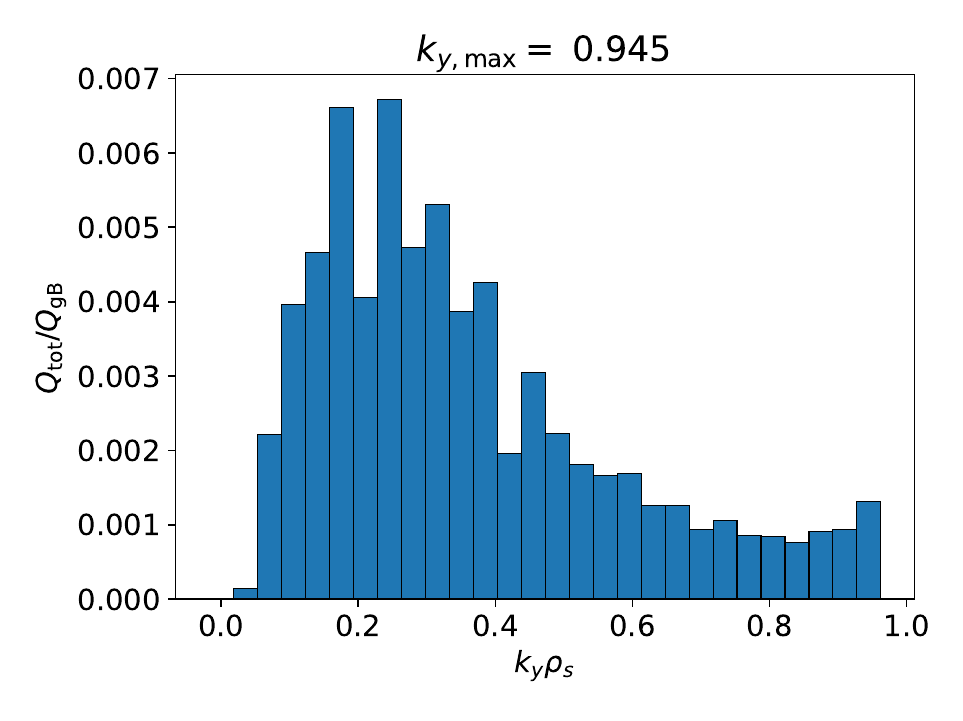}}
    \caption{(a) Saturated value of total, electromagnetic, electrostatic heat fluxes as a function of $k_{y,\mathrm{max}}$. (b) Total heat flux $k_y$ spectrum from the simulation with $k_{y,\mathrm{max}}\rho_s = 0.94$.}
    \label{fig:kymax}
\end{figure}

The value of $k_{y,\mathrm{max}}$ in the nonlinear simulations of the present work is chosen such that the linear $k_y$ spectrum of MTMs is well resolved, while excluding the region in $k_y$ where ETG is the dominant mode.
At $\nu_e=0.42$, MTMs are unstable in the $k_y$ range $0.1 \lesssim k_y\rho_s \lesssim 0.7$, while ETG modes are unstable at $k_y\rho_s\gtrsim 0.9$. This motivated our choice of $k_{y,\mathrm{max}}\rho_s\simeq 0.7$.
We briefly investigate here the effect of $k_{y,\mathrm{max}}$ on the heat flux by comparing the results of four simulations with $k_{y,\mathrm{max}}\rho_s \in \{0.53, 0.67, 0.81, 0.94\}$. We note that the simulation with $k_{y,\mathrm{max}}\rho_s=0.94$ includes a weakly unstable ETG mode. 
The saturated value of heat flux from these simulations is shown in Fig.~\ref{fig:kymax}~(a). The total heat flux depends weakly on $k_{y,\mathrm{max}}$ at $k_{y,\mathrm{max}}\rho_s < 0.9$. On the other hand, the total heat flux increases by a factor of two when $k_{y,\mathrm{max}}\rho_s = 0.94$. 
In this last case, we also note an increase of the electrostatic heat flux, which is driven by the high $k_y$ unstable ETG.

Fig.~\ref{fig:kymax}~(b) shows the total heat flux $k_y$ spectrum from the simulation with $k_{y,\mathrm{max}}\rho_s = 0.94$. Although the heat flux peaks at low $k_y$, significant contribution comes from the high $k_y$ region. In particular, we note that the heat flux decreases monotonically from $k_y\rho_s\simeq 0.3$ to $k_y\rho_s\simeq 0.8$ and increases at $k_y\rho_s>0.8$ due to the ETG contribution. We highlight that only a very small fraction of the unstable ETG spectrum is included in this nonlinear simulation. Multi-scale nonlinear simulations, outside the scope of this work, are required to carefully account for MTM and ETG driven turbulent fluxes.

\section*{References}
\bibliographystyle{unsrt}
\bibliography{bibliography}

\end{document}